\documentclass[twocolumn]{aastex63}

\shorttitle{}
\shortauthors{}
\graphicspath{{./}{figures/}}

\usepackage{amstext}
\usepackage{amsmath}
\usepackage{amssymb}
\usepackage{graphicx}

\newcommand{\aext}{{\bf a}_{\rm ext}}

\newcommand{\beq}{\begin{equation}}
\newcommand{\eeq}{\end{equation}}
\newcommand{\beqar}{\begin{align}}
\newcommand{\eeqar}{\end{align}}

\newcommand{\Ms}{M_\ast}
\newcommand{\Mp}{M_{\rm p}}
\newcommand{\Rs}{R_\ast}
\newcommand{\Rp}{R_{\rm p}}

\begin{document}

\title{Stellar Wind Confinement of Evaporating Exoplanet Atmospheres and its Signatures in 1083~nm Observations}
\correspondingauthor{Morgan MacLeod, Antonija Oklop{\v{c}}i{\'c}}
\email{morgan.macleod@cfa.harvard.edu, a.oklopcic@uva.nl}

\author[0000-0002-1417-8024]{Morgan MacLeod}
\affiliation{Center for Astrophysics $\vert$ Harvard $\&$ Smithsonian 
60 Garden Street, MS-16, Cambridge, MA 02138, USA}

\author[0000-0002-9584-6476]{Antonija Oklop{\v{c}}i{\'c}}
\affil{Anton Pannekoek Institute for Astronomy, University of Amsterdam, Science Park 904, NL-1098 XH Amsterdam, The Netherlands}

\begin{abstract}

Atmospheric escape from close-in exoplanets is thought to be crucial in shaping observed planetary populations. Recently, significant progress has been made in observing this process in action through excess absorption in transit spectra and narrowband light curves. 
We model the escape of initially-homogeneous planetary winds interacting with a stellar wind. The ram pressure balance of the two winds governs this interaction. When the impingement of the stellar wind on the planetary outflow is mild or moderate, the planetary outflow expands nearly spherically through its sonic surface before forming a shocked boundary layer. When the confinement is strong, the planetary outflow is redirected into a cometary tail before it expands to its sonic radius. 
The resultant transmission spectra at the He 1083 nm line are accurately represented by a 1D spherical wind solution in cases of mild to moderate stellar wind interaction. In cases of strong stellar wind interaction, the degree of absorption is enhanced and the cometary tail leads to an extended egress from transit. The crucial features of the wind--wind interaction are, therefore, encapsulated in the light curve of He 1083 nm equivalent width as a function of time. The possibility of extended He 1083 nm absorption well beyond the optical transit carries important implications for planning ``out-of-transit" observations that serve as a baseline for in-transit data. 
\end{abstract}


\section{Introduction} \label{sec:intro}

Atmospheric escape is one of the key physical processes governing the evolution of planets and planetary atmospheres, especially in the case of exoplanets orbiting at close orbital distances from their host stars \citep{Owen2019}. Atmospheric mass loss has been proposed as a possible explanation for the observed lack of intermediate-size planets at short orbital periods (often called the `sub-Jovian desert' or `hot Neptune desert') \citep{SzaboKiss2011, Lundkvist2016} and the gap or valley in the radius distribution of small planets \citep{OwenWu2013, LopezFortney2013, Fulton2017, VanEylen2018}. Improving our understanding of atmospheric escape, especially in the highly irradiated regime that does not exist in our Solar System, is necessary in order to make more robust connections between the properties of mature planetary systems that we can observe in large numbers and various planet formation and evolution scenarios. 

Two main heating mechanisms have been proposed as drivers of atmospheric escape. In the photoevaporation scenario \citep[e.g.][]{Lammer2003, Murray-Clay2009}, the upper layers of a planet's atmosphere absorb extreme ultraviolet and X-ray (XUV) radiation of the host star, forming a thermal pressure gradient which launches a radial outflow from the planet. In the core-powered mass loss model \citep[e.g.][]{Ginzburg2018, GuptaSchlichting2020}, the atmosphere is heated by the stellar bolometric luminosity and the thermal energy stored in the planet's interior during formation. At the moment, either mechanism is able to explain the observed radius valley \citep{Rogers2021}; however, the two processes likely dominate the overall atmospheric mass loss in different regimes of planet and stellar properties, which may help us to eventually discriminate between them \citep{GuptaSchlichting2021}.

To directly observe and characterize atmospheric escape as it occurs, we need to study the properties and kinematics of the high-altitude regions of exoplanetary atmospheres (thermospheres and exospheres). High-resolution transmission spectroscopy at the wavelengths of strong atomic lines, such as the hydrogen Lyman-$\alpha$ \citep[e.g.][]{Vidal-Madjar2003, Lecavelier2010, Ehrenreich2015} or the UV lines of metals \citep[e.g.][]{Linsky2010, BenJaffel2013, Sing2019, GarciaMunoz2021}, allow us to study these low-density environments. Evidence of atmospheric escape can be found either by directly detecting the presence of gas extending beyond the Roche radius of the planet (indicating that the gas is no longer gravitationally bound to the planet) or by probing the gas below the Roche radius and inferring large enough radial velocities for the gas to escape.

The helium line triplet at 1083~nm was recently found to be a good diagnostic of extended and escaping exoplanet atmospheres \citep[e.g.][]{OklopcicHirata2018, Spake2018, Nortmann2018, Allart2018}. Unlike Ly$\alpha$, the helium 1083~nm line does not suffer from significant absorption by the interstellar medium and it can be observed with a fair number of ground-based instruments, making it a much more accessible observing window compared to the UV lines. Consequently, the number of exoplanets with detected excess absorption in the helium 1083~nm line quickly surpassed the number of exoplanets with detected Ly$\alpha$ absorption, although the use of this line as a diagnostic is likely restricted to planets at short orbital distances around stars of spectral type between G and early M \citep{Oklopcic2019}.

Because the first direct evidence of atmospheric escape was obtained through Ly$\alpha$ spectroscopy, most theoretical studies, including those based on 3D hydrodynamic simulations of planetary outflows, have been focused on predicting and interpreting observations in that line \citep[e.g.][]{Bourrier2013a, Villarreal2014, Tripathi2015, Schneiter2016, Carroll-Nellenback2017, McCann2019}. However, similar tools can be used to predict and interpret observations at 1083~nm. The first examples of that approach, focused on modeling individual planets with reported helium detections, have recently been presented by \citet{Allart2018, Khodachenko2021, Shaikhislamov2021, Wang2020a, Wang2020b}. 

In this work, we investigate how interaction with a stellar wind shapes planetary outflows, and how this shaping can be  revealed through observations at 1083~nm. In \S\ref{sec:method} we describe our numerical methods aimed at simulating planetary and stellar winds and their signatures. \S\ref{sec:results} presents the results of our hydrodynamic simulations, which are coupled with a radiative transfer analysis in order to produce synthetic transmission spectra and light curves in the helium 1083 nm line. In \S\ref{sec:discussion}, we discuss the implications of our results for designing optimal observing strategies of planetary transits at 1083~nm and for using those observations to constrain the stellar wind environments of close-in exoplanets. We summarize our results and conclusions in \S\ref{sec:conclusions}.

\section{Method} \label{sec:method}

We model the structure of thermal winds originating from the planet and their interaction with the stellar wind using the three dimensional hydrodynamic simulations developed within the {\tt Athena++} \citep[Version 2019][online at: \url{https://princetonuniversity.github.io/athena}]{2020ApJS..249....4S} code, which is an Eulerian (magneto)hydrodynamic code descended from {\tt Athena} \citep{2008ApJS..178..137S}. We then compute the radiative transfer of stellar light through the simulated planetary wind to produce synthetic absorption spectra at 1083~nm during planet transit. 

We choose to insert hydrodynamic winds into our simulations, instead of self-consistently simulating atmospheric heating and wind launching, in order to remain as agnostic as possible regarding the nature of the wind-driving mechanism(s) and to be able to simulate 3D planetary winds with a broad range of properties in a fast and efficient way that enables searching for the best fit to observational data in future work. This approach also gives us the flexibility to investigate how different wind geometries may affect the observable signatures. In this paper, we study winds that are launched isotropically and leave the investigation of other wind geometries for future work. 

\subsection{Hydrodynamic Simulations}\label{sec:hydromethod}

We solve the equations of inviscid gas dynamics in a frame of reference centered on the host star rotating with the orbital frequency of the planet, $\Omega = \Omega_{\rm orb} = \sqrt{GM/a^3}$, where $M=\Ms + \Mp$, the sum of the star and planet masses. We employ a spherical polar mesh, with origin at the star's center and nested levels of static mesh refinement surrounding the planet. The planet lies along the $-x$-axis at $x=-a$ and the orbital angular momentum is in the $+z$-direction. 

The equations solved are 
\begin{subequations}\label{gaseq}
\begin{align}
\partial_t \rho  + \nabla \cdot \left( \rho {\bf v} \right) &= 0 , \\
\partial_t  \left( \rho {\bf v} \right) + \nabla \cdot \left( \rho {\bf v} {\bf v} + P {\bf I} \right)  &= - \rho \aext , \\
\partial_t E + \nabla \cdot \left[ \left( E+ P \right) {\bf v} \right] &= - \rho \aext \cdot \bf {v} ,
\end{align}
\end{subequations}
expressing  mass continuity, the evolution of gas momenta, and the evolution of gas energies. In the above expressions, $\rho$ is the mass density, $\rho {\bf v}$ is the momentum density, the total energy density is $E = \epsilon + \rho {\bf v} \cdot {\bf v} / 2$,  and $\epsilon$ is the internal energy density. The pressure is $P$, $\bf I$ is the identity tensor, and $\aext$ is the acceleration associated with gravitational forces and the non-inertial frame of reference.  We adopt an ideal gas equation of state, $P=\left(\gamma -1\right) \epsilon$, where $\gamma=1.0001$ is the gas adiabatic index. Thus, along adiabats, gas is nearly isothermal, but wind from the star and planet have different specific entropy. 

The source terms of the star and planet system gravity and the rotating reference frame are contained in the acceleration,
\beq
\aext = -\frac{G \Ms}{|{\bf r} |^3} {\bf r}  -\frac{G \Mp}{|{\bf r_{\rm p}} |^3} {\bf r_{\rm p}} - {\bf \Omega} \times {\bf \Omega} \times {\bf r} - 2 {\bf \Omega} \times {\bf v},
\eeq 
where $\bf r$ is the vector separation from the origin at the stellar center to a given zone, and $\bf r_{\rm p}$ is the vector from the planet center to a  given computational zone center. The vectoral orbital frequency is  ${\bf \Omega} = (0,0,\Omega)$. $\Ms$ and $\Mp$ are the mass of the star and of the planet, respectively. We ignore any gravitational backreaction of the wind distribution on the planetary orbit.

We ignore the influence of 1083~nm radiation pressure on the gas dynamics. In  Appendix \ref{sec:reaction_rates}, we discuss the estimation of the short mean free path in the vicinity of the planet, and how this relates to the solution of surrounding gas dynamics via a fluid, rather than kinetic, approximation. \citet{Wang2020b}'s section 5.1 includes a similar discussion of the relevance of fluid or collisionless treatments of different portions of the planetary outflow, and reaches similar conclusions despite using different values of the particle collision cross section. By contrast, the \citet{Allart2019} model treats the metastable helium as a collisionless particles, and does not capture the redistribution of their momentum to the remainder of the fluid. Because metastable helium is a small mass fraction of the gas (e.g. 1 part in $10^5$ or $10^6$ under typical conditions), we argue that this represents a significant underestimation of the momentum redistribution (and thus, an overestimate of the velocities achieved by the metastable helium outflow). The approach of  \citet{Khodachenko2021} is a multi-fluid model, in which the metastable helium is represented by a separate fluid species than each of the other hydrogen or helium states tracked. \citet{Khodachenko2021} state that separate fluids are coupled by their elastic collisional cross-sections. Their finding is that metastable helium can achieve substantial differential velocity as it is accelerated by the radiation between collisions. This finding is in tension with our and \citet{Wang2020b}'s argument about the prevalence of collusions and the total momentum budget available in the 1083~nm line. This difference is exemplified by tests performed by \citet{Wang2020b} and  \citet{Khodachenko2021}, in which  \citet{Khodachenko2021} found significant effect on the metastable helium kinematics when varying the radiation pressure, while  \citet{Wang2020b} found none. Further investigation is therefore merited, and in particular, there will be tremendous value in methodological approaches that can bridge the fluid and collisionless regimes for simultaneously modeling different lines like metastable helium close to the planet and Lyman $\alpha$ at larger distances. 

The planet and stellar surfaces are allocated to constant density and pressure, related by their respective hydrodynamic escape parameters. Thus, the stellar surface density $\rho_\ast$, and hydrodynamic escape parameter (the ratio of gravitational potential and thermal energy), $\lambda_\ast$, define the stellar surface pressure,
\beq
P_\ast = \rho_\ast \frac{G \Ms}{\gamma \lambda_\ast \Rs }. 
\eeq
The inner boundary of our spherical polar computational domain lies at $r=R_\ast$. The boundary condition is maintained at rest with constant $\rho_\ast$ and $P_\ast$ to establish the stellar wind within the domain.  
Similarly, the planetary boundary is  maintained at constant density $\rho_{\rm p}$ and pressure imposed within $\Rp$ of the planet center, 
\beq
P_{\rm p} = \rho_{\rm p} \frac{G \Mp}{\gamma \lambda_{\rm p} \Rp }. 
\eeq
$\lambda_*$ and $\lambda_{\rm p}$ are free parameters which effectively set the temperatures of the stellar and the planetary wind, respectively.

The planet and the star can rotate with arbitrary frequency. In what follows, we generally assume a non-rotating star and a planet which corotates with the orbital motion, $\Omega_{\rm p} = \Omega$.

The computational domain extends from $r=R_\ast$ to $r=4.5\times10^{12}$~cm (0.3~au), and over the full $4\pi$ in solid angle. The angular coordinate ranges are $0 < \theta < \pi$ and $0< \phi < 2\pi$. The base mesh is composed of $8\times6\times12$ meshblocks of $16^3$ zones. Zones are spaced logarithmically in $r$ and evenly in $\theta$ and $\phi$ such that near-cubic zone shapes are maintained throughout the volume to the maximum extent possible. Near the poles, we reduce the number of effective zones in the $\phi$ direction to avoid zones with an extreme aspect ratio. This operation is performed by averaging conserved quantities across these zones, rather than re-arranging the mesh structure itself \citep{2018ApJ...863....5M,2018ApJ...868..136M}. We utilize $N_{\rm SMR}=5$ additional levels of static mesh refinement around the planet. The maximally refined region lies within $10~\Rp$ of the planet center.

\subsection{Radiative Transfer and Synthetic Spectra}\label{sec:spectramethod}

We assume the gas  throughout the computational domain has solar composition, with hydrogen, helium, and metal mass fractions of $X=0.738$, $Y=0.248$, $Z=0.014$, and that, in steady state, detailed balance is achieved in each cell of the simulation grid. For hydrogen atoms, the recombination rate equals the photoionization rate (ionization via collisions is assumed to be negligible at temperatures expected in planetary atmospheres):
\begin{equation}
    n_e n_{\mathrm{H^+}} \alpha_\mathrm{rec} = n_{\mathrm{H^0}}\phi e^{-\tau} \ \mbox{,}
\end{equation}
where $n_e$, $n_{\mathrm{H^0}}$, and $n_{\mathrm{H^+}}$ are the number densities of free electrons, neutral hydrogen, and ionized hydrogen, respectively, and $\alpha_\mathrm{rec} =  2.59\times 10^{-13} (T/10^4)^{-0.7}$~cm$^3$~s$^{-1}$ is the case-B hydrogen recombination rate at temperature $T$ \citep{OsterbrockFerland, Tripathi2015}. The hydrogen photoioniziation rate in each cell is given by $\Phi e^{-\tau}$. Unattenuated photoionization rate $\Phi$  is calculated from the input stellar flux\footnote{We use the empirical flux of a K6V-type star HD 85512 from the MUSCLES survey \citep{France2016}.} $F_\nu$ as
\begin{equation}
\phi = \int_{\nu_0}^\infty \frac{F_{\nu}}{h\nu}a_\nu   
\label{eq:phi}
\end{equation}
where $\nu_0$ is the frequency corresponding to 13.6~eV photons and the hydrogen photoionization cross section as function of frequency ($\nu$) is given by \citep{OsterbrockFerland}:
\begin{equation}
a_\nu = 6.3\times 10^{-18} \frac{\exp{\left(4-\frac{4\tan^{-1}\delta}{\delta}\right)}}{1-\exp{(-2\pi/\delta)}} \left(\frac{\nu_0}{\nu}\right)^4 \ \mbox{[cm$^2$]},
\end{equation}
where $\delta = \sqrt{\nu/\nu_0 -1}$. We iteratively calculate the optical depth to hydrogen-ionizing photons for each cell in the simulation grid, starting from an initial assumption of having 10\% of hydrogen in the ionized state. Assuming that hydrogen ionization is the main source of free electrons (i.e. ignoring the electrons produced by the ionization of helium or metals), we calculate the number density of free electrons in each subsequent iteration as:
\begin{equation}
    n_e = n_\mathrm{H^+} = \frac{\phi e^{-\tau_0}}{2\alpha_\mathrm{rec}} \left[ \sqrt{1 +\frac{4\rho X \alpha_\mathrm{rec}}{\phi e^{-\tau_0}m_H}} - 1\right] \ \mbox{,}
\end{equation}
where $\tau_0$ is the cumulative optical depth between the star and the grid location, and $m_H$ is the mass of the hydrogen atom.

The number density of helium atoms is given by $n_\mathrm{He} = \frac{\rho Y}{m_\mathrm{He}}$. We calculate the the fractions of helium atoms in the helium ground (singlet) state ($f_1$) and the metastable 2$^3$S (triplet) state ($f_3$) by taking into account the rates of various processes that cause mixing between these two neutral states and the (singly) ionized state:
\begin{itemize}
    \item photoionization of the ground state, $f_1 \phi_1 e^{-\tau_1}$ \citep[with photoionization cross sections from ][]{Brown1971}, and of the metastable state, $f_3 \phi_3 e^{-\tau_3}$ \citep{Norcross1971}, where $\tau_1$ and $\tau_3$ are the optical depths to radiation ionizing the ground and the metastable state, respectively,
    \item recombination into the ground state, $(1-f_1-f_3)n_e \alpha_1$, and into the metastable state, $(1-f_1-f_3)n_e \alpha_3$, with temperature-dependent recombination rate coefficients from \citet{OsterbrockFerland}, 
    \item radiative decay between the metastable state and the ground state, $f_3 A_{31}$ \citep{Drake1971}, 
    \item collisions with free electrons, $q_{13}$ and $q_{31}$ (with temperature-dependent rate coefficients from \citet{Bray2000}), and with neutral hydrogen atoms, $Q_{31}$ \citep{RobergeDalgarno1982}.
\end{itemize}  
Detailed balance sets the rate of processes which populate a given state equal to the rate of processes which depopulate that state. For the helium ground state and the metastable state, this reads: 
\begin{eqnarray}
 (1-f_1-f_3)n_e \alpha_1 & +& f_3A_{31} + f_3 (n_e q_{31} + n_\mathrm{H^0}Q_{31}) \\
\nonumber &= & f_1 \phi_1 e^{-\tau_1} + f_1 n_e q_{13}\\
(1-f_1-f_3)n_e \alpha_3 &+& f_1 n_e q_{13}\\
\nonumber &=& f_3 A_{31} + f_3 \phi_3 e^{-\tau_3} +  f_3 (n_e q_{31} + n_\mathrm{H^0}Q_{31}) \ \mbox{.}
\end{eqnarray}
After a few ($\sim 5$) iterations, we obtain the density of helium atoms in the metastable state ($n_\mathrm{He}f_3$) throughout the simulation domain.The relative rates of different processes (de)populating the excited helium state in different parts of our computational domain are presented in \autoref{sec:reaction_rates}.  We note that this treatment is most appropriate for the planetary, rather than the stellar wind. In the stellar wind, where the density is significantly lower and velocities are higher, advection plays an important role in setting the level populations. In our models, we find that stellar wind material contributes negligibly to the eventual absorption signature, and we elect to ignore the detailed treatment of that material. 

For the transfer of photoionizing radiation, we treat the planet as transparent. This has the effect of ignoring the shadow of the planet itself, and only including the cumulative optical depth of the gaseous wind. This choice is equivalent to assuming that material advecting through the shadow region maintains its ionization state, rather than recombining to an ionization equilibrium with its instantaneous location. Models including coupled radiation and hydrodynamics can better treat this region, and test the validity of a simple approximation like this. For example, \citet{Wang2020b}'s figure 2 shows the effect of the shadow region in a model of WASP-107b. In their model, the recombination time of shadow material is generally similar to or longer than the dynamical timescale of flow near the planet, motivating our approximation. Finally, we emphasize that the shadow region is largely unobservable because it is blocked by the planet itself during the optical transit, implying that the treatment of this material should not have a significant effect on the transmission spectra.

To obtain a synthetic spectrum around the helium line triplet at 1083 nm, we choose the location of our ``observer" and calculate the integrated optical depth, as function of wavelength, along rays, $l$, between the star and the distant observer:
\begin{equation}
    \tau (\lambda) =  \sum_{i=1}^{3}  \int n_\mathrm{He}f_3  \sigma_i \Phi (\lambda) dl \ \mbox{,}
\end{equation}
where the absorption cross sections for the three components of the triplet are $\sigma_1 =0.0080$ cm$^2$~Hz, $\sigma_2 =0.0048 $ cm$^2$~Hz, $\sigma_3 = 0.0016$ cm$^2$~Hz. We assume a Voigt line profile, $\Phi(\lambda)$, with a Gaussian component which depends on the local gas temperature, $T=\gamma \frac{ \mu m_H }{k_B} \frac{P}{\rho}$. The wavelength offset from the line center takes into account the inertial-frame gas velocity projected onto the line-of-sight. Quantities are interpolated from the simulation mesh via nearest-neighbor interpolation to the zone centers, such that any sampling point within a zone volume is assigned the quantity of that zone. Rays intercepting the planet interior are assigned infinite optical depth such that they are completely blocked at all wavelengths. 

The rays are weighted based on the quadratic limb darkening law and combined to produce the average transmission spectrum $F_\mathrm{in}/F_\mathrm{out} (\lambda) = e^{-\overline{\tau} (\lambda)}$. We allocate rays to more densely cover the region near the planet, and weight them according to their relative area. The convergence and numerical parameters of our ray tracing are tested and documented in Appendix \ref{sec:numerical_param}.

\subsection{Model Parameters and Simulations}

The models used in this paper use parameters similar to the properties of WASP-107, the first system in which helium has been detected \citep{Spake2018}. We adopt $M_\ast = 0.68$~M$_{\odot}= 1.36\times10^{33}$~g, $R_\ast=0.67$~R$_{\odot}=4.67\times10^{10}$~cm, and $M_{\rm p} = 0.096$~M$_\mathrm{Jupiter}=1.82\times10^{29}$~g, $R_{\rm p}=0.94$~R$_\mathrm{Jupiter}=6.71\times10^{9}$~cm \citep{Piaulet2021, Anderson2017}.  We specify the planetary boundary condition with $\lambda_{\rm p}=5$ in each case. 
The degree of planetary rotation is set to be locked with the orbital frequency,  $\Omega_{\rm p} = \Omega_{\rm orb}$.
The planet density is chosen so that the time-averaged planetary mass-loss rate is similar across cases A, B, and C. This implies that $\rho_{\rm p}$ is higher in case C than in B or A. 
We vary the stellar boundary condition in order to change the stellar mass-loss rate. In each case, we adopt $\lambda_\ast=15$. These variations result in a stellar mass-loss rate that increases approximately an order of magnitude between models. The rate of model C is most similar to the solar wind mass-loss rate of $2\times 10^{12}$~g~s$^{-1}$. Our models A--C utilize 5 nested levels of static mesh refinement around the planet. For convergence studies, we additionally run variations of model A with $N_{\rm SMR}$=3, 4 and 6 (A3, A4, A6, respectively). Each of our models is run for $3\times10^6$s, or slightly more than 6 orbits of the planet about the star. We find that our key quantities of interest reach a quasi-steady state after about one orbit or $\sim 5\times10^5$~s. 

Table \ref{simtable} summarizes our simulations and the emergent model parameters. These include  the time average stellar and planetary mass loss rates ($\langle \dot M_\ast \rangle$ and $\langle \dot M_{\rm p} \rangle$) measured from $10^6$~s until the end of the simulation. We also list the parameters to Gaussian fits to model light curves, which are described in more detail in Section \ref{sec:results}.

\begin{table*}
\begin{center}
\begin{tabular}{ccccccccccc}
\hline
model & $N_{\rm SMR}$ & $\rho_\ast$ & $P_\ast$ & $\rho_{\rm p}$ & $P_{\rm p}$ & $\langle\dot M_\ast\rangle$ & $\langle\dot M_{\rm p}\rangle$ & $g_\mu$ & $g_a$ & $g_\sigma$ \\
\hline
 & & [g~cm$^{-3}$] & [erg~cm$^{-3}$] &  [g~cm$^{-3}$] & [erg~cm$^{-3}$] & [g~s$^{-1}$] & [g~s$^{-1}$] & [hr] & [\AA] & [hr]  \\ 
\hline
A & 5 & $8.31\times10^{-16}$ & $1.07\times10^{-1}$ & $1.55\times10^{-16}$ & $5.63\times10^{-5}$ & $1.71\times10^{10}$ & $9.10\times10^{9}$ & -0.018 & 0.031 & 0.85 \\
B & 5 & $8.31\times10^{-15}$ & $1.07\times10^0$ & $1.55\times10^{-16}$ & $5.63\times10^{-5}$ & $1.78\times10^{11}$ & $8.70\times10^{9}$ & -0.064 & 0.033 & 0.93 \\
C & 5 &  $8.31\times10^{-14}$ & $1.07\times10^{1}$ & $2.92\times10^{-16}$ & $1.06\times10^{-4}$  & $1.78\times10^{12}$ & $7.24\times10^{9}$ & 0.32 & 0.093 & 0.97 \\
\hline
A3 & 3 & $8.31\times10^{-16}$ & $1.07\times10^{-1}$ & $1.55\times10^{-16}$ & $5.63\times10^{-5}$ & $2.10\times10^{10}$ & $1.16\times10^{10}$ & -- & -- & --  \\
A4 & 4 & $8.31\times10^{-16}$ & $1.07\times10^{-1}$ & $1.55\times10^{-16}$ & $5.63\times10^{-5}$ & $2.11\times10^{10}$ & $1.20\times10^{10}$  & -- & -- & --  \\
A6 & 6 & $8.31\times10^{-16}$ & $1.07\times10^{-1}$ & $1.55\times10^{-16}$ & $5.63\times10^{-5}$ & $2.13\times10^{10}$ & $1.14\times10^{10}$  & -- & -- & --  \\
\hline
\end{tabular}
\caption{Table of simulation parameters and results. Our models are grouped into three cases, models A, B, C. In addition, several variations of model A are run with different resolution surrounding the planet, as parameterized by $N_{\rm SMR}$. Note that because the planetary outflow is suppressed by the stellar wind confinement in case C, model C has higher density and pressure at the planet boundary condition in order to achieve a similar time-averaged mass loss rate as models A and B. Finally, as a baseline, we compare models A--C to a spherical model ``Sph," which is a 1D isothermal (Parker) wind model with the same mass loss rate as model A.}\label{simtable}
\end{center}
\end{table*}

\section{Results} \label{sec:results}

\begin{figure*}[tbp]
\begin{center}
\includegraphics[width=\textwidth]{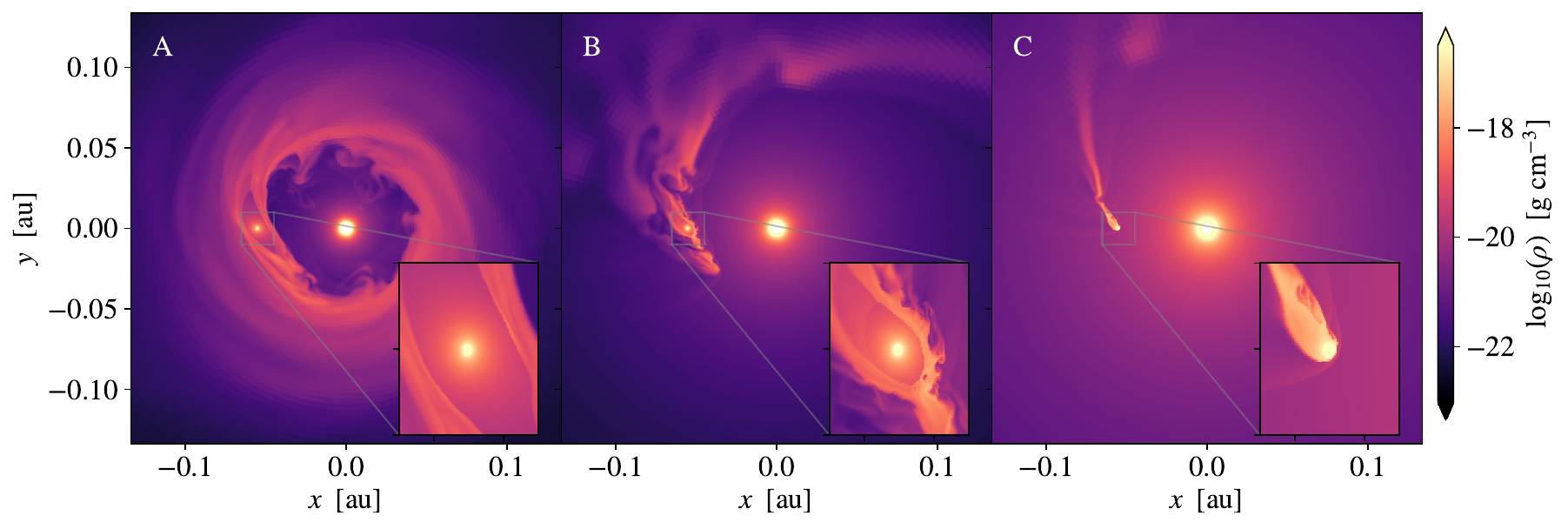}
\caption{Slices of the logarithm of gas density in the orbital midplane, showing hydrodynamic winds from model exoplanets interacting with stellar winds of varying magnitude. In case A, with the lowest stellar mass loss rate, the planetary outflow forms a torus around the star, while in case C it is focused into a dense, cometary tail. Cases A and B contain a cavity of freely expanding planetary wind (shown in the zoom-in panels), while in Case C, the boundary shock penetrates to the planet. }
\label{fig:global}
\end{center}
\end{figure*}

In this section, we examine transmission spectra and light curves of the helium 1083~nm triplet based on our models with varying stellar winds. The nature of stellar wind and planet wind interactions has been the subject of a number of recent hydrodynamic and magnetohydrodynamic simulation efforts \citep[e.g.][]{StoneProga2009,Cohen2011,Bisikalo2013,Matsakos2015,Christie2016,Shaikhislamov2016,Carroll-Nellenback2017,Villarreal2018,Bharati2019,DaleyYates2019,Debrecht2019,Khodachenko2019,McCann2019,Wang2020a,Wang2020b,Harbach2020,Carolan2021, Villarreal2021}. As we will highlight, several characteristic regimes of global morphology emerge as the relative ram pressure $(\rho v^2)$  of these winds is varied \citep[for example, see figure 13 of][]{Matsakos2015}. Winds that begin subsonic near the surfaces of both objects become supersonic as they expand. The wind interaction region is therefore defined by bow shocks that are shaped by the ram pressures of the superimposed winds. 

\subsection{Wind Shaping}

We begin, in Figure \ref{fig:global}, by examining the global density structure of the interacting star and planet winds in the orbital plane in our simulations. The star is located at the origin of the coordinate system, while the planet lies along the $-x$-axis, and is highlighted in the zoomed-in region. 

Three very different global morphologies are seen in Figure \ref{fig:global} as the stellar wind mass-loss rate is varied by a factor of 100 from case A to C. In case A, and at even lower stellar mass-loss rates, we observe a cavity of stellar wind which is confined in the equator by escaping planetary material. The stellar wind finds a free path to escape through a bipolar outflow perpendicular to the orbital plane. At lower stellar mass-loss rates, we observe very similar global morphology with slightly changing extent of the central region of unshocked stellar wind. 
By contrast, as the stellar wind mass-loss rate and momentum increase, the planetary wind is swept into an increasingly comet-like tail. A region of unshocked planetary wind exists in models A and B, but disappears in model C as the wind--wind interaction region is compressed to near the planetary surface. 

\citet{McCann2019} classifies the interaction scenarios seen in our Figure \ref{fig:global} as cases of ``weak'', ``intermediate", and ``strong" confinement of the planet wind by the stellar wind (see, for example, their figures 10 and 11). When the stellar wind compresses the planetary wind boundary to below its sonic surface, we see the morphology shown in Figure C, where subsonic outflow is redirected around the planet toward the tail. This also leads to a suppression of the outflow rate given certain planetary surface conditions  \citep{Christie2016,Vidotto2020,Wang2020b,Carolan2021}. In the example of case C, we increase the planetary surface density by a factor of approximately 2 in order to achieve a similar outflow rate. However, in this context it is worth highlighting the result of previous authors that the relationship between surface conditions and loss rate is nonlinear; a model with the same planetary surface conditions as models A and B lead to a planetary mass-loss rate that was suppressed by a factor of $\sim 4$ in the environment of the case C stellar wind. 

Because ram pressure balance dictates the location and morphology of the wind--wind interaction, a useful order of magnitude comparison can be to compare the ram pressure of the two winds at their respective sonic points, $\rho_\ast c_{\rm s, \ast}^2$ and $\rho_{\rm p} c_{\rm s,p}^2$. In the context of an isothermal wind, the sonic point density is approximately $\rho(r_{\rm s}) \approx \rho_{\rm b} \exp(1.5-\lambda)$ where $\rho_{\rm b}$ represents the base density at the object surface. Given this, we can express the sonic point ram pressure ratio as 
\beq
\label{Pram}
\frac{\rho_{\rm p} c_{\rm s,p}^2}{\rho_\ast c_{\rm s, \ast}^2} \approx \left( \frac{\rho_{\rm p}}{\rho_\ast} \right)  \left( \frac{M_{\rm p}}{M_\ast} \right) \left( \frac{R\ast}{R_{\rm p}} \right)  \left( \frac{\lambda_\ast}{\lambda_{\rm p}} \right) \exp\left( \lambda_\ast - \lambda_{\rm p} \right).
\eeq
Computing this ratio for the models A, B, and C, we find $( \rho_{\rm p} c_{\rm s,p}^2) /( \rho_\ast c_{\rm s, \ast}^2 ) \approx 11.5$, $1.15$, and $2.16$, respectively. In general, larger values of this ratio imply that the planetary wind penetrates further into the stellar wind environment before reaching its location of ram pressure balance. In cases A, and B, these ratios provide useful guidance. In case A, the ram pressure of the planetary outflow clearly exceeds that of the stellar wind, which allows it to form a torus about the star. In case B, the ram pressures are similar. In case C, the result of this simple ratio is somewhat misleading, because the planetary outflow is suppressed by the stellar wind interaction (such that the wind density is less than predicted by the analytic expression above given the planetary density), and because the planetary outflow never reaches its sonic point -- thus the applicability of this ratio has limitations when $( \rho_{\rm p} c_{\rm s,p}^2) /( \rho_\ast c_{\rm s, \ast}^2 ) \lesssim 1$. A more complete, albeit numerical, approach would involve comparing the velocity and density profiles of the two winds in isolation and finding the location of ram pressure balance along the ray between the planet and star.

\begin{figure*}[tbp]
\begin{center}
\includegraphics[width=\textwidth]{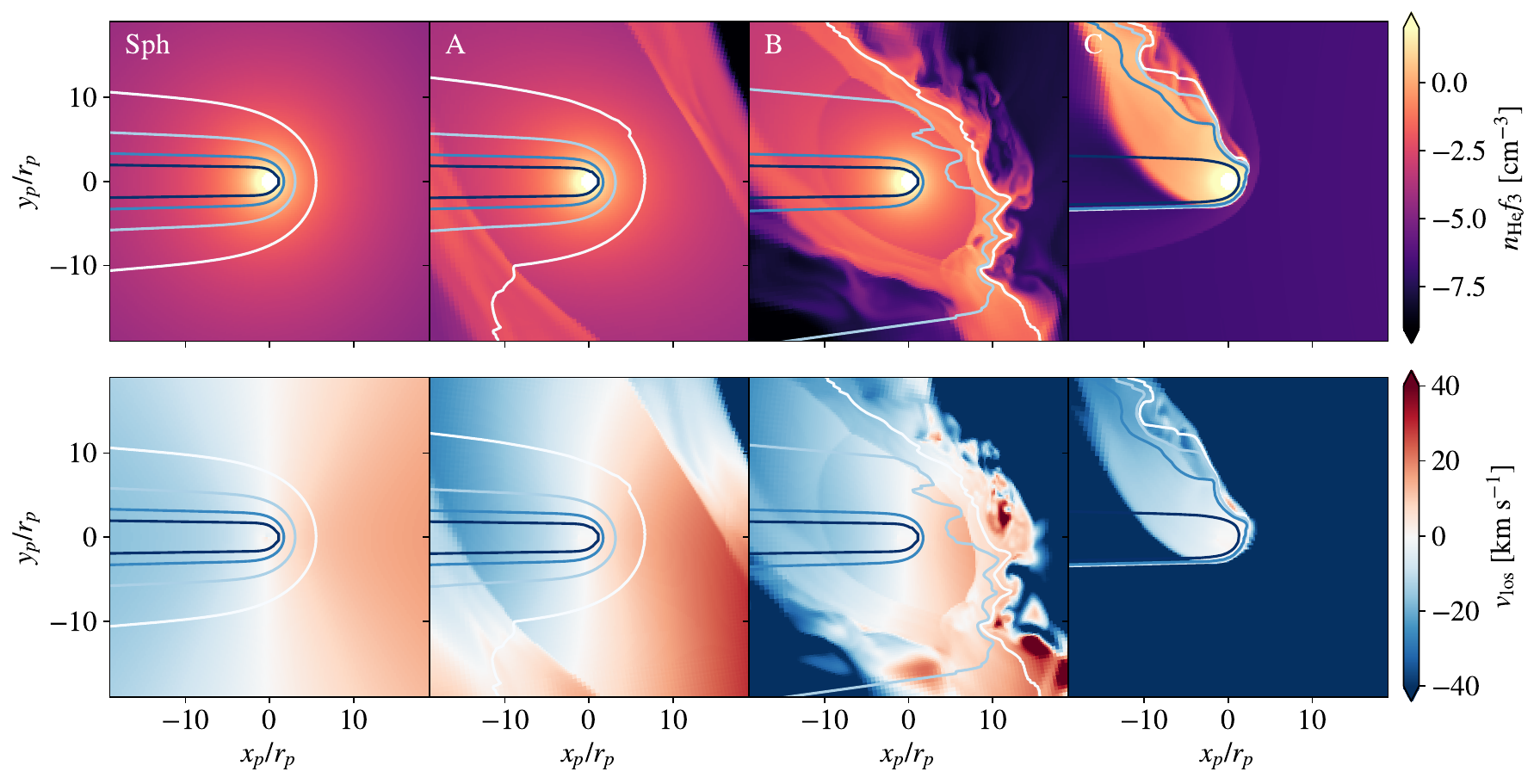}
\caption{Metastable helium number density (top), and line of sight velocity (bottom). Contours mark the line-of-sight optical depth in the strong (red) component of the helium 1083~nm line of $\tau=10^{-3}$, $10^{-2}$, $10^{-1}$, and $1$ from light to dark. In this image the star is in the $+x_p$-direction and the observer is in the $-x_p$-direction. The flow density, kinematics, and optical depth in the immediate vicinity of the planet combine to form the excess absorption signatures at 1083 nm. Here we observe that with increasing stellar wind strength, the asymmetry of the region around the planet also increases. While in case A the $\tau$ surfaces are relatively symmetric  and similar to the spherical model ``Sph", in cases B and C, increasing departures from symmetry become apparent. We note that in the highly-confined case C, nearly all of the planetary outflow is blueshifted and lags behind the planet in orbital phase.  }
\label{fig:rt}
\end{center}
\end{figure*}

In Figure \ref{fig:rt}, we zoom in on the region surrounding the planet  and compare models A--C with a spherical Parker wind (labeled ``Sph"). The upper panels display the number density of metastable helium, $n_{\rm He} f_3$, as determined by the iterative radiative transfer post-processing discussed in Section \ref{sec:spectramethod}.  Contours in Figure \ref{fig:rt} display the cumulative optical depth from the star at the center of the primary doublet of the 1083~nm line, with contours representing $\tau=10^{-3}$, $10^{-2}$, $10^{-1}$, and $1$ from light to dark. The contours nicely illustrate why the helium line is a powerful probe of extended and escaping planetary atmospheres: observable levels of absorption form in the region spanning several planetary radii around the planet (and in some cases reaching close to or even beyond the Roche radius). 

The progression of the ``Sph" model to cases A--C in Figure \ref{fig:rt} reveals increasing asymmetry in spatial distribution of metastable helium around the planet. While in case A, the near-planet flow and contours of optical depth are very similar to the spherical outflow model, in case C they diverge dramatically.  In particular, we highlight the increasing degree of leading versus trailing asymmetry that arises as the stellar wind increasingly shapes and compresses the planet wind. In case C, where the planetary material is compressed into a narrow, cometary tail, the $\tau=0.1$ surface extends to more than 15 planetary radii in the $+y$-direction (trailing the planet along its orbital motion). This implies that the center of the opaque region in the metastable helium line lags behind the planet's optical transit, creating a significant difference between early-transit and late-transit transmission spectra, and resulting in a prolonged helium egress, as observed in transits of WASP-69b \citep{Nortmann2018} and WASP-107b \citep{Spake2021}.

Examining the line-of-sight velocities in the lower panels of Figure \ref{fig:rt} is revealing because it shows how Doppler-shifted material adds together to form the composite transmission spectrum. At the characteristic wavelength of the helium triplet line, a 10~km~s$^{-1}$ line-of-sight velocity corresponds to a 0.36~\AA\ wavelength shift. Some of the important features in this flow in cases A and B are the overall pattern of nearly radial expansion from the planet in the unshocked planet wind region. This is seen most clearly in redshifted material lying at $+x_p$ positions, while the majority of blueshifted material lies at $-x_p$ positions. In case A, the pattern of radial velocity shows how orbital motion distorts and confines this region. By contrast, in case C, the wind never expands through its sonic point, and instead is redirected toward the trailing surface of the planet while subsonic. Rather than boundary layers between the winds, we see that nearly all of the planetary outflow is blueshifted, and swept up in the compressed tail.

In cases A--C, we observe hydrodynamic instabilities along the wind--wind interfaces in Figures \ref{fig:global} and \ref{fig:rt}. In case A, Raleigh-Taylor instabilities are especially prominent as denser planetary material overlies the lower-density but faster-moving stellar wind. This instability feeds plumes that penetrate into the central region and fall toward the star \citep{DaleyYates2019}. In cases B and C, we see Kelvin-Helmholtz shear instabilities develop along the wind--wind boundary layers \citep[e.g.][]{Christie2016}. We note that while these unstable interfaces are physical, the exact nature of the turbulence that is excited is affected by finite spatial resolution and numerical diffusion in our models.  

\subsection{Transmission Spectra}

\begin{figure*}[tbp]
\begin{center}
\includegraphics[width=\textwidth]{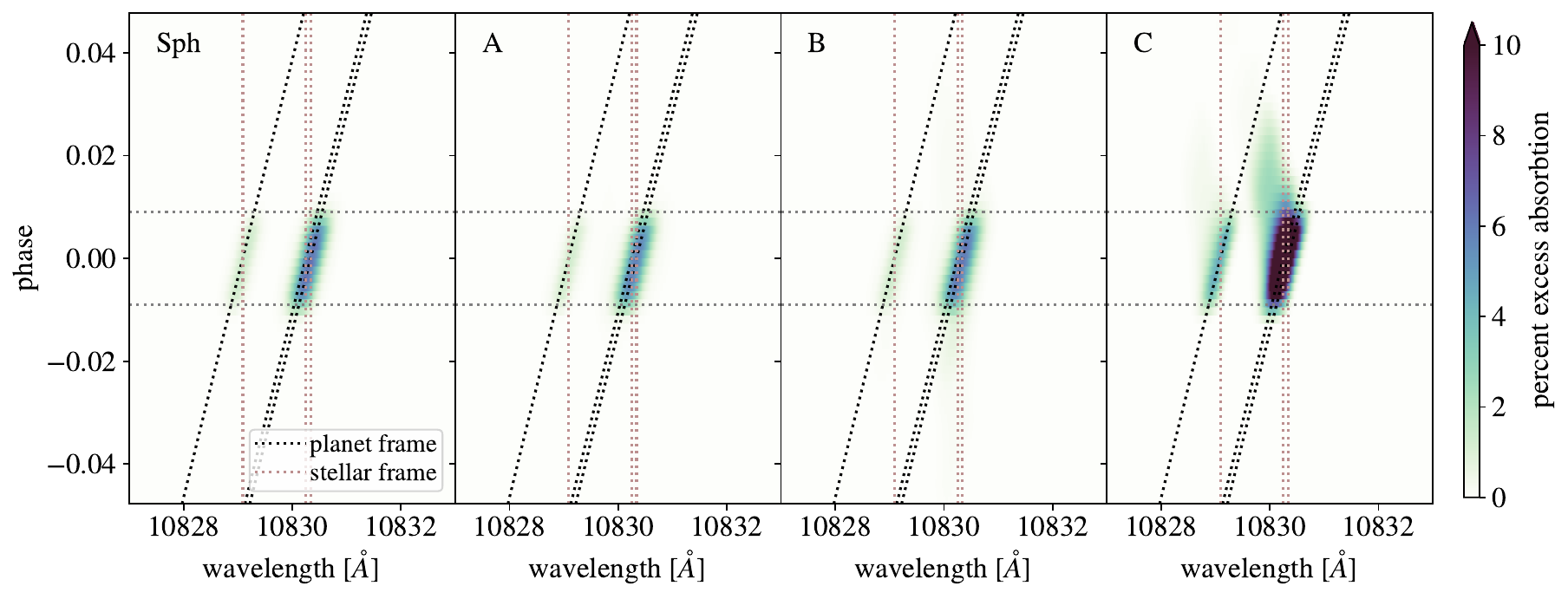}
\includegraphics[width=\textwidth]{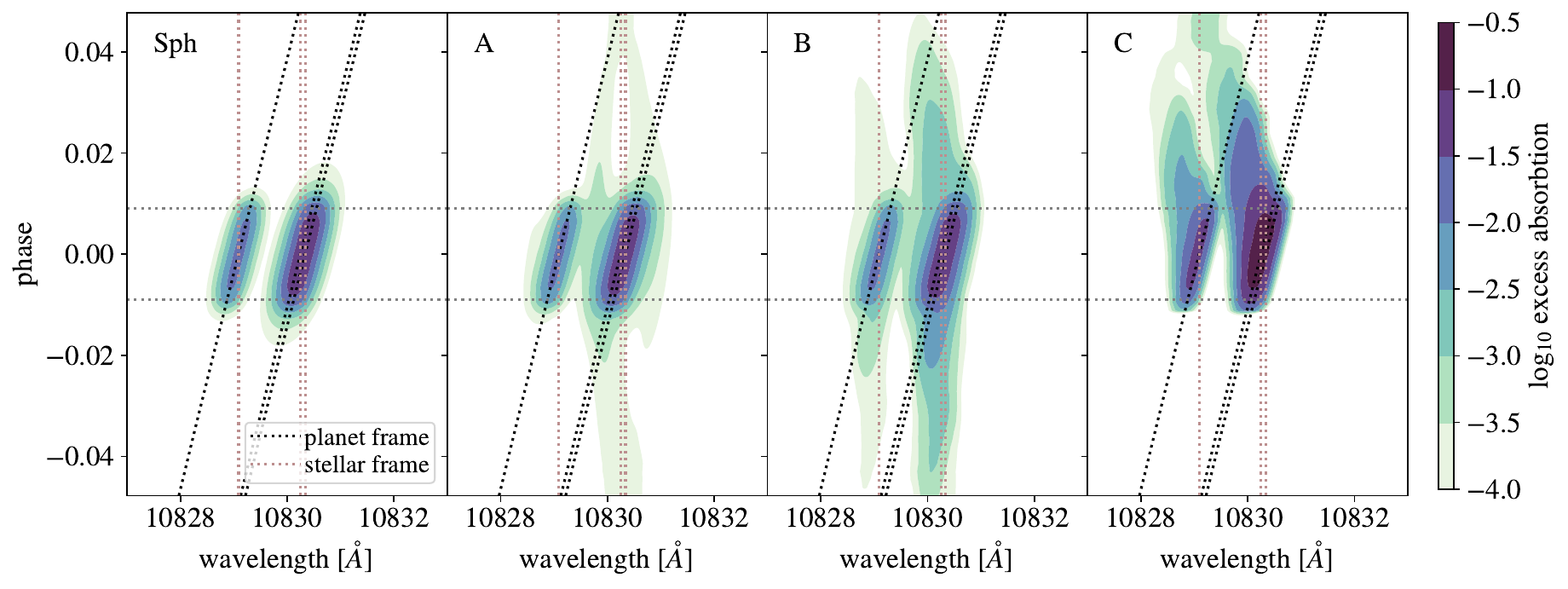}
\caption{Excess absorption of metastable helium in the stellar frame. Horizontal lines show the extent of the optical transit, while the slanted lines trace the Doppler-shifted wavelengths of the helium triplet in the planet frame. The upper and lower panels are identical except for using a linear or logarithmic scale. We see that there is a significant contribution to absorption outside of the optical transit in cases B and C. While in case B, there is absorption both before and after transit, in case C, the major portion of the excess absorption comes after optical transit. The kinematic structures of blueshifted and redshifted layers, especially outside of the optical transit, trace the wind--wind interaction regions.    }
\label{fig:abs}
\end{center}
\end{figure*}

Figure \ref{fig:abs} shows synthetic transmission spectra, plotted as excess absorption as a function of wavelength in the stellar frame. These spectra are computed via radiative transfer post-processing of our simulation snapshots. For this figure, we post-process a single model snapshot at different angles to represent time advancing through the transit. We assume that there is no inclination of the planetary orbital plane relative to the observer. In this case, the optical (broadband) transit lasts between $-0.01$ and $0.01$ in phase or $\sim 2.5$~hr in total. However, Figure \ref{fig:abs} shows phases significantly before and after the optical transit itself in order to explore the effects of extended planetary wind material along the orbital path. Angled lines show the Doppler-shifted planet frame wavelengths of the helium triplet. 

Differing intensities and morphologies of the spectral phase data are immediately clear in Figure \ref{fig:abs}. Secondly, helium triplet absorption extends well before and after the optical transit itself, especially in cases B and C. The spherical, case A, and case B models show the bulk of their excess absorption tracing the planet in velocity space, and similar overall intensities. These features indicate that the bulk of the absorbing material is in the immediate vicinity of the planet and comoving along its orbit. Connecting back to the spatial analysis of Figure \ref{fig:rt}, we can understand the origin of these features in nearly-spherical outflow from the planet in the first few planet radii. Thus, in cases of mild and moderate stellar wind confinement, the absorbtion is similar to that of the spherically symmetric model based on a 1D Parker wind profile \citep{OklopcicHirata2018}. 
By contrast, the strong stellar wind confinement of case C is revealed through the increased transit depth, and the asymettry of the excess absorption relative to the planet frame. We see material lagging in orbital phase and largely blueshifted. 

Examining the logarithmically-scaled transmission spectra, we see that there are kinematic traces of the degree of wind--wind interaction that extend well outside the phases of optical transit. The origin of these features is in dense regions of shocked planetary outflow, rather than the initial scale heights surrounding the planet. Cases A and B show excess absorption nearly at rest in the stellar frame both before and after optical transit, while in case C, the strength of the stellar wind implies that interacting material is further blueshifted and only trails the planet in phase. Reconstructing this level of detail observationally is clearly challenging, but there is a high density of information available in these reconstructions, particularly when contextualized by the simulation models, like in Figure \ref{fig:rt}.

\subsection{Narrow-band Light Curves}

\begin{figure*}[tbp]
\begin{center}
\includegraphics[width=0.60\textwidth]{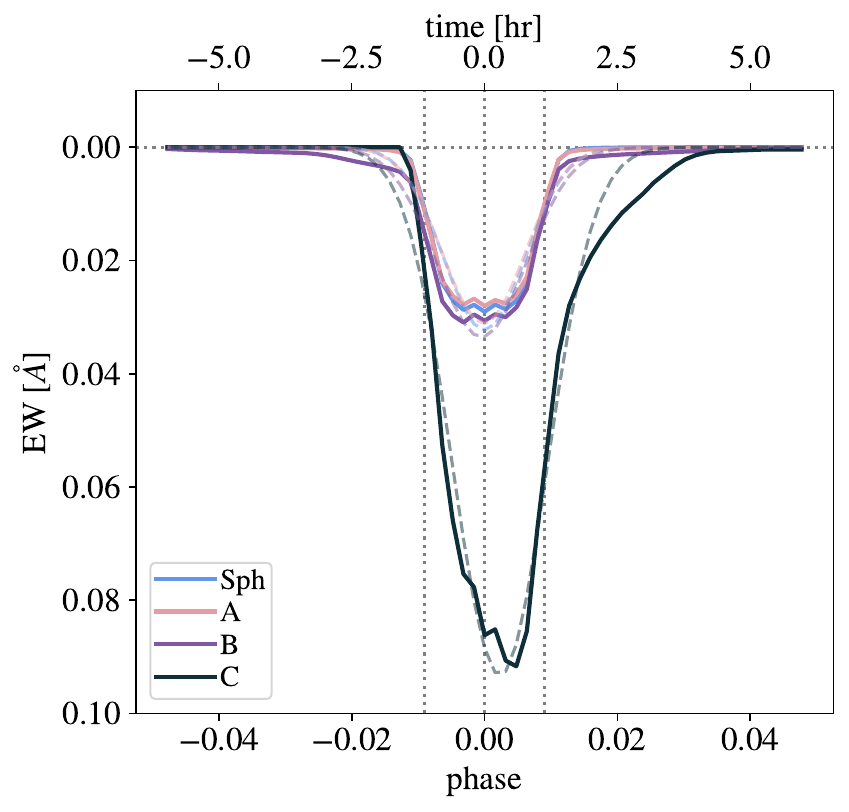}
\includegraphics[width=0.38\textwidth]{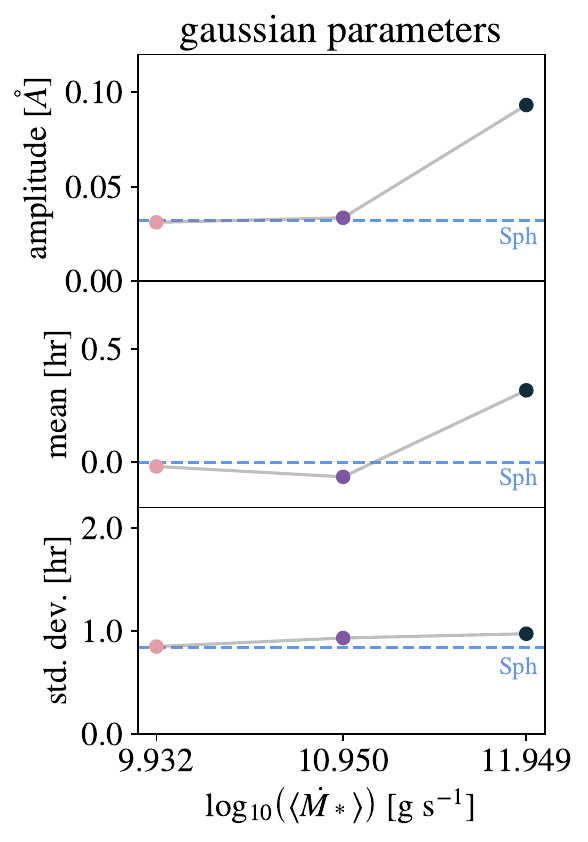}
\caption{Light curve of excess equivelent width of metastable helium absorption as a function of time relative to mid transit,  as well as light curve parameters of gaussian fits to the light curve morphologies. We compare models A--C to the spherical model, ``Sph." In cases A, and B, the light curve morphology and depth closely trace that of the spherical model. Most of the absorption originates in the nearly-spherical outflow near the planet. In case C, confinement by the stellar wind dramatically increases the equivalent width, and shifts the time of maximal absorption relative to the optical transit.}
\label{fig:lc}
\end{center}
\end{figure*}

In Figure \ref{fig:lc}, we turn our attention to the total excess absorption in the metastable helium triplet line as measured by the line equivalent width. Although this quantity can be derived from high-spectral resolution observations, it is particularly relevant in the context of upcoming medium-resolution observations with the \textit{James Webb Space Telescope}, as well as ground-based photometric observations using a beam-shaping diffuser and an ultra-narrowband filter centered on the wavelength of the helium triplet, recently presented by \citet{Vissapragada2020} and \citet{Paragas2021}.

In Figure \ref{fig:lc}, we compute the equivalent-width light curve for each of our model transits.  Recently, \citet{Wang2020a,Wang2020b} have presented similar light curve analyses for their models of WASP-69b and WASP-107b, respectively. Dashed lines in Figure \ref{fig:lc} represent a best-fit Gaussian to the equivalent width as a function of time. In the right hand panels, we plot the parameters of this best-fit Gaussian as a function of the time-averaged stellar mass loss rate, comparing to the spherical planetary outflow case -- which includes no stellar wind component -- with horizontal lines. 

 With increasing $\dot M_\ast$ from model A to C, we see that the light curves of equivalent width change both amplitude and shape. While we have discussed some of these differences in the context of the spectral phase curves of Figure \ref{fig:abs}, it is striking how clear they are in the EW light curve. The EW light curve of cases A and B is extremely similar to that of the spherical model. At higher $\dot M_\ast$, in case C, the maximum EW is larger, indicating deeper absorption in the helium triplet line. Physically, this occurs because increasing confinement by the ram pressure of the stellar wind concentrates the planetary outflow, increasing $\tau$ (as shown in Figure \ref{fig:rt}). Thus, for a given outflow rate, the planetary wind material in case C is higher density. Indeed, because of the supression of the outflow by the stellar wind, in order to obtain a similar time-averaged planetary mass loss rate for case C, we increase the base  density of the planetary atmosphere in model C, as noted in Table \ref{simtable}. We also observe that light curves become increasingly asymmetric relative to mid transit with increasing distortion of the planet's evaporative wind by the stellar wind, which is particularly notable in case C. Related asymmetries are visible in the parameter variations of \citet{Wang2020b}'s Figure 8, and are discussed in the context of Lyman $\alpha$ absorption in their Figure 12 and previously in e.g. \citet{McCann2019}.  

 Fitting a Gaussian to the light curve encapsulates some aspects of the changing equivalent width as a function of time. For example, the dependence of the peak equivalent width on $\dot M_\ast$ is traced by the amplitude parameter, while the shift of the peak equivalent width to after mid transit is captured by the mean. However, other aspects of the light curves are clearly informative but are not traced by the relatively crude metric of a Gaussian fit. For example, the abrupt onset and extended wing of absorption in case C are not well described by the Gaussian fit. These results suggest that studying the shape of the ingress and egress in metastable helium equivalent width can be quite revealing with respect to the size and nature of the metastable helium absorbing region, including the extent to which it shaped by the stellar wind.

\section{Discussion}\label{sec:discussion}

\subsection{Implications for Constraining Stellar Wind Environments }
At face value, the strength of the stellar wind is not directly probed by measurements of excess absorption at the helium 1083~nm lines. In particular, excess metastable helium absorption occurs primarily in the region within a few planet radii -- which is, of course, what makes this line a powerful diagnostic of the launching of photoevaporative flows. A priori, we might not expect that global properties would imprint on these regions in the immediate vicinity of the planetary atmosphere to an observable extent.  And yet, the stellar wind environment dramatically informs the shaping of the flow evaporating from the planet \citep[Figures 1 and 2, and][]{Carroll-Nellenback2017,Debrecht2019,DaleyYates2019,McCann2019,Wang2020a,Wang2020b}. It may also affect the degree of atmospheric escape itself \citep[e.g.][]{Christie2016,Vidotto2020,Carolan2021}. As we have shown in the light curves of Figure \ref{fig:lc}, in addition to extending excess metastable helium absorption to well outside the optical transit, stellar wind interaction strongly increases the amplitude of helium triplet absorption by shocking and confining the escaping planet wind. 

While our model systems are inspired by WASP-107b, we caution that the models presented in this paper are intended to be illustrative rather than representative of a particular system.  In particular, the model parameters are chosen to span differing regimes of stellar wind ram pressure relative to planet wind ram pressure.  As previous work has argued, this ram pressure ratio is the controlling dimensionless parameter with respect to the hydrodynamic morphology of the wind--wind interaction \citep[e.g.][]{Matsakos2015,Carolan2021}.   Our results demonstrate that when the ram pressure of the stellar wind causes it to impinge only mildly or moderately on the planetary wind (as in cases A and B), spherical outflow models provide accurate representations of the observable metastable helium equivalent widths and kinematics. Only when the planetary wind is strongly confined, as in case C, does a significant departure from the spherical solution result, accompanied by significant enhancements of line depth at a given planetary mass loss rate. 

Disentangling the ram pressure ratio of stellar to planetary winds may be observationally tractable in some cases. In particular, if signatures of extended ingress and egress or dramatic phase asymmetry are not present, it is likely reasonable to conclude that stellar wind interaction is mild or moderate and is not significantly increasing the effective line depth through confinement. In these cases, spherical models of planetary outflows \citep[e.g.][]{OklopcicHirata2018, Lampon2020, dosSantos2021} provide accurate estimates of planetary mass loss rates. By contrast, when an extended egress is observed, this is a strong signature of stellar wind confinement of the planetary outflow. This is accompanied by an enhancement in the line equivalent width at a given planetary mass loss rate. An open-source framework for spherical wind analysis ``p-winds" was recently released by \citet{LDSpwind}; our results highlight the value of these simplified models in all but the most dramatic cases of stellar wind interaction. 

Finally, it is worth highlighting that the hydrodynamic interaction between stellar and planetary winds depends on these materials being in the fluid-like collisional regime, where the typical mean free path is much less than the flow scale. We discuss the estimation of these quantities in Appendix \ref{sec:reaction_rates}. In the collisionless regime, we might expect different species to not couple as strongly, and radiation pressure on species like metastable helium or Lyman-$\alpha$ could contribute significantly to the shaping of the planetary outflow  \citep{Allart2018,Allart2019,Debrecht2020,Khodachenko2021}.

\subsection{Implications for Observing Strategies}

Figures \ref{fig:abs} and \ref{fig:lc} show that helium transits may  extend for several hours before or after the optical transit. As a result, the observed ``out-of-transit" spectra taken immediately before or after the broadband transit may not actually be representative of the uncontaminated stellar spectrum, since they may contain contributions from an extended planetary tail (either from the leading or trailing arm), as demonstrated recently by \citet{Spake2021}. Consequently, the ``measured" transit depth in the helium line could be underestimated. Case B, shown in Figure \ref{fig:lc}, is particularly interesting because the time evolution of the helium absorption EW exhibits a ``plateau" on either side of the optical transit. This type of light curve might be difficult to discern observationally from an uncontaminated light curve, if the observing window is limited to a relatively narrow range in phase around the transit midpoint, which is typically the case in ground-based observations. Space telescopes would be able to provide long enough time coverage to characterize the full light curve, and \textit{James Webb Space Telescope/NIRSpec} would be the optimal choice for that experiment, since \textit{Hubble Space Telescope/WFC3} might not be able to detect the small excess absorption hours before or after the optical transit due to its limited spectral resolution.

\section{Summary \& Conclusion}\label{sec:conclusions}

In this paper we present a hydrodynamic simulation and radiative transfer post-processing methodology for computing model spectra of the metastable helium line at 1083~nm in transiting exoplanets. Our hydrodynamic simulations include parameterized treatments of the planetary and stellar winds, with the goal of using this flexibility to understand how the changes in wind properties impact the observable properties of evaporating atmospheres. This sort of parameterized modeling is complemented and informed by models which attempt self-consistent treatment of the full radiation-(magneto)hydrodynamic problem of photoionization radiation heating the planet atmosphere and launching outflows \citep[e.g.][]{Tripathi2015,Wang2018,McCann2019,Wang2020a,Wang2020b,Khodachenko2021}.  

We have used these models to study the interaction of a planetary wind  with stellar winds of varying mass-loss rates. Interaction with the stellar wind is crucial in defining the global geometry of the planetary outflow \citep[e.g.][]{Matsakos2015,Carroll-Nellenback2017}, but is generally not thought to be strongly constrained by transmission spectra of Lyman-$\alpha$ or metastable helium alone. In this study, we emphasize that the stellar wind impacts the observable properties of metastable helium transits in concrete ways. Some key findings of our work are:
\begin{enumerate}  

    \item At low and moderate levels of stellar-wind impingement, as inferred from the planet wind to stellar wind ram pressure balance, equation \eqref{Pram}, a homogeneous planetary outflow remains roughly spherical close to the planet. In transit spectra and equivalent width light curves, these cases are accurately modeled by spherically symmetric planetary winds. Substantial departures from the spherical solutions occur when the planetary flow is strongly confined into a cometary tail (as in case C). 
    
    \item Many crucial aspects of the relative ram pressure of star and planet winds are encapsulated in light curves of equivalent width as a function of time. Excess absorption in these light curves is asymmetric and extends beyond the transit egress in cases of strong confinement of the planetary outflow by the stellar wind.
    
    \item The prolonged duration of excess helium absorption well outside of broadband transit, in cases of moderate and strong wind-wind interaction, implies that observing strategies need to accommodate taking a baseline ``stellar" spectrum as far from the phase of transit as possible. 

\end{enumerate}

The need to characterize stellar spectra as far from transit as possible may complicate planning for transit observations. However, the potential payoff is that measuring the absolute depth of excess absorption and the detailed shape of the light curve, including any excess absorption before and after broadband transit, provides information about both the planetary wind and the stellar wind environment surrounding an evaporating planet. This stellar wind environment modifies the expected signatures of metastable helium absorption in important ways. For example, changes in amplitude due to stellar wind confinement could be confused with a larger planetary mass loss rate. However, our results suggest that with further development, a method that combines narrowband light curves with spectra before, during, and after optical transit has sufficient information to break degeneracies that exist in mid-transit spectral analysis alone.

\acknowledgments
We gratefully acknowledge helpful discussions with  A. Dupree, L. Hillenbrand, J. Kirk, D. Linssen, A. Loeb, J. McCann, F. Nail, J. Owen, and J. Spake.
This work was supported by the National Science Foundation under Grant No. 1909203. 
Resources supporting this work were provided by the NASA High-End Computing (HEC) Program through the NASA Advanced Supercomputing (NAS) Division at Ames Research Center. 
This work used the Extreme Science and Engineering Discovery Environment (XSEDE), which is supported by National Science Foundation grant number ACI-1548562. In particular, use of XSEDE resource Stampede2 at TACC through allocation TG-AST200014 enabled this work. AO gratefully acknowledges support from the Dutch Research Council NWO Veni grant.

\vspace{5mm}
\software{IPython \citep{PER-GRA:2007}; SciPy \citep{2020SciPy-NMeth};  NumPy \citep{van2011numpy};  matplotlib \citep{Hunter:2007}; Astropy \citep{2013A&A...558A..33A}; Athena++ \citep{2020ApJS..249....4S}; XSEDE \citep{xsede}}

\clearpage

\begin{thebibliography}{}
\expandafter\ifx\csname natexlab\endcsname\relax\def\natexlab#1{#1}\fi
\providecommand{\url}[1]{\href{#1}{#1}}
\providecommand{\dodoi}[1]{doi:~\href{http://doi.org/#1}{\nolinkurl{#1}}}
\providecommand{\doeprint}[1]{\href{http://ascl.net/#1}{\nolinkurl{http://ascl.net/#1}}}
\providecommand{\doarXiv}[1]{\href{https://arxiv.org/abs/#1}{\nolinkurl{https://arxiv.org/abs/#1}}}

\bibitem[{{Allart} {et~al.}(2018){Allart}, {Bourrier}, {Lovis}, {Ehrenreich},
  {Spake}, {Wyttenbach}, {Pino}, {Pepe}, {Sing}, \& {Lecavelier des
  Etangs}}]{Allart2018}
{Allart}, R., {Bourrier}, V., {Lovis}, C., {et~al.} 2018, Science, 362, 1384,
  \dodoi{10.1126/science.aat5879}

\bibitem[{{Allart} {et~al.}(2019){Allart}, {Bourrier}, {Lovis}, {Ehrenreich},
  {Aceituno}, {Guijarro}, {Pepe}, {Sing}, {Spake}, \&
  {Wyttenbach}}]{Allart2019}
---. 2019, arXiv e-prints, arXiv:1901.08073.
\newblock \doarXiv{1901.08073}

\bibitem[{{Anderson} {et~al.}(2017){Anderson}, {Collier Cameron}, {Delrez},
  {Doyle}, {Gillon}, {Hellier}, {Jehin}, {Lendl}, {Maxted}, {Madhusudhan},
  {Pepe}, {Pollacco}, {Queloz}, {S{\'e}gransan}, {Smalley}, {Smith}, {Triaud},
  {Turner}, {Udry}, \& {West}}]{Anderson2017}
{Anderson}, D.~R., {Collier Cameron}, A., {Delrez}, L., {et~al.} 2017, \aap,
  604, A110, \dodoi{10.1051/0004-6361/201730439}

\bibitem[{{Astropy Collaboration} {et~al.}(2013){Astropy Collaboration},
  {Robitaille}, {Tollerud}, {Greenfield}, {Droettboom}, {Bray}, {Aldcroft},
  {Davis}, {Ginsburg}, {Price-Whelan}, {Kerzendorf}, {Conley}, {Crighton},
  {Barbary}, {Muna}, {Ferguson}, {Grollier}, {Parikh}, {Nair}, {Unther},
  {Deil}, {Woillez}, {Conseil}, {Kramer}, {Turner}, {Singer}, {Fox}, {Weaver},
  {Zabalza}, {Edwards}, {Azalee Bostroem}, {Burke}, {Casey}, {Crawford},
  {Dencheva}, {Ely}, {Jenness}, {Labrie}, {Lim}, {Pierfederici}, {Pontzen},
  {Ptak}, {Refsdal}, {Servillat}, \& {Streicher}}]{2013A&A...558A..33A}
{Astropy Collaboration}, {Robitaille}, T.~P., {Tollerud}, E.~J., {et~al.} 2013,
  \aap, 558, A33, \dodoi{10.1051/0004-6361/201322068}

\bibitem[{{Ben-Jaffel} \& {Ballester}(2013)}]{BenJaffel2013}
{Ben-Jaffel}, L., \& {Ballester}, G.~E. 2013, \aap, 553, A52,
  \dodoi{10.1051/0004-6361/201221014}

\bibitem[{{Bharati Das} {et~al.}(2019){Bharati Das}, {Basak}, {Nandy}, \&
  {Vaidya}}]{Bharati2019}
{Bharati Das}, S., {Basak}, A., {Nandy}, D., \& {Vaidya}, B. 2019, \apj, 877,
  80, \dodoi{10.3847/1538-4357/ab18ad}

\bibitem[{{Bisikalo} {et~al.}(2013){Bisikalo}, {Kaygorodov}, {Ionov},
  {Shematovich}, {Lammer}, \& {Fossati}}]{Bisikalo2013}
{Bisikalo}, D., {Kaygorodov}, P., {Ionov}, D., {et~al.} 2013, \apj, 764, 19,
  \dodoi{10.1088/0004-637X/764/1/19}

\bibitem[{{Bourrier} {et~al.}(2013){Bourrier}, {Lecavelier des Etangs},
  {Dupuy}, {Ehrenreich}, {Vidal-Madjar}, {H{\'e}brard}, {Ballester},
  {D{\'e}sert}, {Ferlet}, {Sing}, \& {Wheatley}}]{Bourrier2013a}
{Bourrier}, V., {Lecavelier des Etangs}, A., {Dupuy}, H., {et~al.} 2013, \aap,
  551, A63, \dodoi{10.1051/0004-6361/201220533}

\bibitem[{{Bray} {et~al.}(2000){Bray}, {Burgess}, {Fursa}, \&
  {Tully}}]{Bray2000}
{Bray}, I., {Burgess}, A., {Fursa}, D.~V., \& {Tully}, J.~A. 2000, \aaps, 146,
  481, \dodoi{10.1051/aas:2000277}

\bibitem[{{Brown}(1971)}]{Brown1971}
{Brown}, R.~L. 1971, \apj, 164, 387, \dodoi{10.1086/150851}

\bibitem[{{Carolan} {et~al.}(2021){Carolan}, {Vidotto}, {Villarreal D'Angelo},
  \& {Hazra}}]{Carolan2021}
{Carolan}, S., {Vidotto}, A.~A., {Villarreal D'Angelo}, C., \& {Hazra}, G.
  2021, \mnras, 500, 3382, \dodoi{10.1093/mnras/staa3431}

\bibitem[{{Carroll-Nellenback} {et~al.}(2017){Carroll-Nellenback}, {Frank},
  {Liu}, {Quillen}, {Blackman}, \& {Dobbs-Dixon}}]{Carroll-Nellenback2017}
{Carroll-Nellenback}, J., {Frank}, A., {Liu}, B., {et~al.} 2017, \mnras, 466,
  2458, \dodoi{10.1093/mnras/stw3307}

\bibitem[{{Christie} {et~al.}(2016){Christie}, {Arras}, \& {Li}}]{Christie2016}
{Christie}, D., {Arras}, P., \& {Li}, Z.-Y. 2016, \apj, 820, 3,
  \dodoi{10.3847/0004-637X/820/1/3}

\bibitem[{{Cohen} {et~al.}(2011){Cohen}, {Kashyap}, {Drake}, {Sokolov}, \&
  {Gombosi}}]{Cohen2011}
{Cohen}, O., {Kashyap}, V.~L., {Drake}, J.~J., {Sokolov}, I.~V., \& {Gombosi},
  T.~I. 2011, \apj, 738, 166, \dodoi{10.1088/0004-637X/738/2/166}

\bibitem[{{Daley-Yates} \& {Stevens}(2019)}]{DaleyYates2019}
{Daley-Yates}, S., \& {Stevens}, I.~R. 2019, \mnras, 483, 2600,
  \dodoi{10.1093/mnras/sty3310}

\bibitem[{{Debrecht} {et~al.}(2019){Debrecht}, {Carroll-Nellenback}, {Frank},
  {Blackman}, {Fossati}, {McCann}, \& {Murray-Clay}}]{Debrecht2019}
{Debrecht}, A., {Carroll-Nellenback}, J., {Frank}, A., {et~al.} 2019, arXiv
  e-prints, arXiv:1906.00075.
\newblock \doarXiv{1906.00075}

\bibitem[{{Debrecht} {et~al.}(2020){Debrecht}, {Carroll-Nellenback}, {Frank},
  {Blackman}, {Fossati}, {McCann}, \& {Murray-Clay}}]{Debrecht2020}
---. 2020, \mnras, 493, 1292, \dodoi{10.1093/mnras/staa351}

\bibitem[{{Dos Santos} {et~al.}(2021{\natexlab{a}}){Dos Santos}, {Vidotto},
  {Vissapragada}, {Alam}, {Allart}, {Bourrier}, {Kirk}, {Seidel}, \&
  {Ehrenreich}}]{dosSantos2021}
{Dos Santos}, L.~A., {Vidotto}, A.~A., {Vissapragada}, S., {et~al.}
  2021{\natexlab{a}}, arXiv e-prints, arXiv:2111.11370.
\newblock \doarXiv{2111.11370}

\bibitem[{{Dos Santos} {et~al.}(2021{\natexlab{b}}){Dos Santos}, {Vidotto},
  {Vissapragada}, {Alam}, {Allart}, {Bourrier}, {Kirk}, {Seidel}, \&
  {Ehrenreich}}]{LDSpwind}
---. 2021{\natexlab{b}}, arXiv e-prints, arXiv:2111.11370.
\newblock \doarXiv{2111.11370}

\bibitem[{{Drake}(1971)}]{Drake1971}
{Drake}, G.~W. 1971, \pra, 3, 908, \dodoi{10.1103/PhysRevA.3.908}

\bibitem[{{Ehrenreich} {et~al.}(2015){Ehrenreich}, {Bourrier}, {Wheatley},
  {Lecavelier des Etangs}, {H{\'e}brard}, {Udry}, {Bonfils}, {Delfosse},
  {D{\'e}sert}, {Sing}, \& {Vidal-Madjar}}]{Ehrenreich2015}
{Ehrenreich}, D., {Bourrier}, V., {Wheatley}, P.~J., {et~al.} 2015, \nat, 522,
  459, \dodoi{10.1038/nature14501}

\bibitem[{{France} {et~al.}(2016){France}, {Parke Loyd}, {Youngblood}, {Brown},
  {Schneider}, {Hawley}, {Froning}, {Linsky}, {Roberge}, {Buccino},
  {Davenport}, {Fontenla}, {Kaltenegger}, {Kowalski}, {Mauas}, {Miguel},
  {Redfield}, {Rugheimer}, {Tian}, {Vieytes}, {Walkowicz}, \&
  {Weisenburger}}]{France2016}
{France}, K., {Parke Loyd}, R.~O., {Youngblood}, A., {et~al.} 2016, \apj, 820,
  89, \dodoi{10.3847/0004-637X/820/2/89}

\bibitem[{{Fulton} {et~al.}(2017){Fulton}, {Petigura}, {Howard}, {Isaacson},
  {Marcy}, {Cargile}, {Hebb}, {Weiss}, {Johnson}, {Morton}, {Sinukoff},
  {Crossfield}, \& {Hirsch}}]{Fulton2017}
{Fulton}, B.~J., {Petigura}, E.~A., {Howard}, A.~W., {et~al.} 2017, \aj, 154,
  109, \dodoi{10.3847/1538-3881/aa80eb}

\bibitem[{{Garc{\'\i}a Mu{\~n}oz} {et~al.}(2021){Garc{\'\i}a Mu{\~n}oz},
  {Fossati}, {Youngblood}, {Nettelmann}, {Gandolfi}, {Cabrera}, \&
  {Rauer}}]{GarciaMunoz2021}
{Garc{\'\i}a Mu{\~n}oz}, A., {Fossati}, L., {Youngblood}, A., {et~al.} 2021,
  \apjl, 907, L36, \dodoi{10.3847/2041-8213/abd9b8}

\bibitem[{{Ginzburg} {et~al.}(2018){Ginzburg}, {Schlichting}, \&
  {Sari}}]{Ginzburg2018}
{Ginzburg}, S., {Schlichting}, H.~E., \& {Sari}, R. 2018, \mnras, 476, 759,
  \dodoi{10.1093/mnras/sty290}

\bibitem[{{Gupta} \& {Schlichting}(2020)}]{GuptaSchlichting2020}
{Gupta}, A., \& {Schlichting}, H.~E. 2020, \mnras, 493, 792,
  \dodoi{10.1093/mnras/staa315}

\bibitem[{{Gupta} \& {Schlichting}(2021)}]{GuptaSchlichting2021}
---. 2021, \mnras, 504, 4634, \dodoi{10.1093/mnras/stab1128}

\bibitem[{{Harbach} {et~al.}(2020){Harbach}, {Moschou}, {Garraffo}, {Drake},
  {Alvarado-G{\'o}mez}, {Cohen}, \& {Fraschetti}}]{Harbach2020}
{Harbach}, L.~M., {Moschou}, S.~P., {Garraffo}, C., {et~al.} 2020, arXiv
  e-prints, arXiv:2012.05922.
\newblock \doarXiv{2012.05922}

\bibitem[{Hunter(2007)}]{Hunter:2007}
Hunter, J.~D. 2007, Computing In Science \& Engineering, 9, 90

\bibitem[{{Khodachenko} {et~al.}(2021){Khodachenko}, {Shaikhislamov},
  {Fossati}, {Lammer}, {Rumenskikh}, {Berezutsky}, {Miroshnichenko}, \&
  {Efimof}}]{Khodachenko2021}
{Khodachenko}, M.~L., {Shaikhislamov}, I.~F., {Fossati}, L., {et~al.} 2021,
  \mnras, 503, L23, \dodoi{10.1093/mnrasl/slab015}

\bibitem[{{Khodachenko} {et~al.}(2019){Khodachenko}, {Shaikhislamov}, {Lammer},
  {Berezutsky}, {Miroshnichenko}, {Rumenskikh}, {Kislyakova}, \&
  {Dwivedi}}]{Khodachenko2019}
{Khodachenko}, M.~L., {Shaikhislamov}, I.~F., {Lammer}, H., {et~al.} 2019,
  \apj, 885, 67, \dodoi{10.3847/1538-4357/ab46a4}

\bibitem[{{Lammer} {et~al.}(2003){Lammer}, {Selsis}, {Ribas}, {Guinan},
  {Bauer}, \& {Weiss}}]{Lammer2003}
{Lammer}, H., {Selsis}, F., {Ribas}, I., {et~al.} 2003, \apjl, 598, L121,
  \dodoi{10.1086/380815}

\bibitem[{{Lamp{\'o}n} {et~al.}(2020){Lamp{\'o}n}, {L{\'o}pez-Puertas}, {Lara},
  {S{\'a}nchez-L{\'o}pez}, {Salz}, {Czesla}, {Sanz-Forcada}, {Molaverdikhani},
  {Alonso-Floriano}, {Nortmann}, {Caballero}, {Bauer}, {Pall{\'e}}, {Montes},
  {Quirrenbach}, {Nagel}, {Ribas}, {Reiners}, \& {Amado}}]{Lampon2020}
{Lamp{\'o}n}, M., {L{\'o}pez-Puertas}, M., {Lara}, L.~M., {et~al.} 2020, \aap,
  636, A13, \dodoi{10.1051/0004-6361/201937175}

\bibitem[{{Lecavelier Des Etangs} {et~al.}(2010){Lecavelier Des Etangs},
  {Ehrenreich}, {Vidal-Madjar}, {Ballester}, {D{\'e}sert}, {Ferlet},
  {H{\'e}brard}, {Sing}, {Tchakoumegni}, \& {Udry}}]{Lecavelier2010}
{Lecavelier Des Etangs}, A., {Ehrenreich}, D., {Vidal-Madjar}, A., {et~al.}
  2010, \aap, 514, A72, \dodoi{10.1051/0004-6361/200913347}

\bibitem[{{Linsky} {et~al.}(2010){Linsky}, {Yang}, {France}, {Froning},
  {Green}, {Stocke}, \& {Osterman}}]{Linsky2010}
{Linsky}, J.~L., {Yang}, H., {France}, K., {et~al.} 2010, \apj, 717, 1291,
  \dodoi{10.1088/0004-637X/717/2/1291}

\bibitem[{{Lopez} \& {Fortney}(2013)}]{LopezFortney2013}
{Lopez}, E.~D., \& {Fortney}, J.~J. 2013, \apj, 776, 2,
  \dodoi{10.1088/0004-637X/776/1/2}

\bibitem[{{Lundkvist} {et~al.}(2016){Lundkvist}, {Kjeldsen}, {Albrecht},
  {Davies}, {Basu}, {Huber}, {Justesen}, {Karoff}, {Silva Aguirre}, {van
  Eylen}, {Vang}, {Arentoft}, {Barclay}, {Bedding}, {Campante}, {Chaplin},
  {Christensen-Dalsgaard}, {Elsworth}, {Gilliland}, {Handberg}, {Hekker},
  {Kawaler}, {Lund}, {Metcalfe}, {Miglio}, {Rowe}, {Stello}, {Tingley}, \&
  {White}}]{Lundkvist2016}
{Lundkvist}, M.~S., {Kjeldsen}, H., {Albrecht}, S., {et~al.} 2016, Nature
  Communications, 7, 11201, \dodoi{10.1038/ncomms11201}

\bibitem[{{MacLeod} {et~al.}(2018{\natexlab{a}}){MacLeod}, {Ostriker}, \&
  {Stone}}]{2018ApJ...863....5M}
{MacLeod}, M., {Ostriker}, E.~C., \& {Stone}, J.~M. 2018{\natexlab{a}}, \apj,
  863, 5, \dodoi{10.3847/1538-4357/aacf08}

\bibitem[{{MacLeod} {et~al.}(2018{\natexlab{b}}){MacLeod}, {Ostriker}, \&
  {Stone}}]{2018ApJ...868..136M}
---. 2018{\natexlab{b}}, \apj, 868, 136, \dodoi{10.3847/1538-4357/aae9eb}

\bibitem[{{Matsakos} {et~al.}(2015){Matsakos}, {Uribe}, \&
  {K{\"o}nigl}}]{Matsakos2015}
{Matsakos}, T., {Uribe}, A., \& {K{\"o}nigl}, A. 2015, \aap, 578, A6,
  \dodoi{10.1051/0004-6361/201425593}

\bibitem[{{McCann} {et~al.}(2019){McCann}, {Murray-Clay}, {Kratter}, \&
  {Krumholz}}]{McCann2019}
{McCann}, J., {Murray-Clay}, R.~A., {Kratter}, K., \& {Krumholz}, M.~R. 2019,
  \apj, 873, 89, \dodoi{10.3847/1538-4357/ab05b8}

\bibitem[{{Murray-Clay} {et~al.}(2009){Murray-Clay}, {Chiang}, \&
  {Murray}}]{Murray-Clay2009}
{Murray-Clay}, R.~A., {Chiang}, E.~I., \& {Murray}, N. 2009, \apj, 693, 23,
  \dodoi{10.1088/0004-637X/693/1/23}

\bibitem[{{Norcross}(1971)}]{Norcross1971}
{Norcross}, D.~W. 1971, Journal of Physics B Atomic Molecular Physics, 4, 652,
  \dodoi{10.1088/0022-3700/4/5/006}

\bibitem[{{Nortmann} {et~al.}(2018){Nortmann}, {Pall{\'e}}, {Salz},
  {Sanz-Forcada}, {Nagel}, {Alonso-Floriano}, {Czesla}, {Yan}, {Chen},
  {Snellen}, {Zechmeister}, {Schmitt}, {L{\'o}pez-Puertas}, {Casasayas-Barris},
  {Bauer}, {Amado}, {Caballero}, {Dreizler}, {Henning}, {Lamp{\'o}n}, {Montes},
  {Molaverdikhani}, {Quirrenbach}, {Reiners}, {Ribas}, {S{\'a}nchez-L{\'o}pez},
  {Schneider}, \& {Zapatero Osorio}}]{Nortmann2018}
{Nortmann}, L., {Pall{\'e}}, E., {Salz}, M., {et~al.} 2018, Science, 362, 1388,
  \dodoi{10.1126/science.aat5348}

\bibitem[{{Oklop{\v{c}}i{\'c}}(2019)}]{Oklopcic2019}
{Oklop{\v{c}}i{\'c}}, A. 2019, \apj, 881, 133, \dodoi{10.3847/1538-4357/ab2f7f}

\bibitem[{{Oklop{\v{c}}i{\'c}} \& {Hirata}(2018)}]{OklopcicHirata2018}
{Oklop{\v{c}}i{\'c}}, A., \& {Hirata}, C.~M. 2018, \apj, 855, L11,
  \dodoi{10.3847/2041-8213/aaada9}

\bibitem[{{Osterbrock} \& {Ferland}(2006)}]{OsterbrockFerland}
{Osterbrock}, D.~E., \& {Ferland}, G.~J. 2006, {Astrophysics of gaseous nebulae
  and active galactic nuclei}

\bibitem[{{Owen}(2019)}]{Owen2019}
{Owen}, J.~E. 2019, Annual Review of Earth and Planetary Sciences, 47, 67,
  \dodoi{10.1146/annurev-earth-053018-060246}

\bibitem[{{Owen} \& {Wu}(2013)}]{OwenWu2013}
{Owen}, J.~E., \& {Wu}, Y. 2013, \apj, 775, 105,
  \dodoi{10.1088/0004-637X/775/2/105}

\bibitem[{{Paragas} {et~al.}(2021){Paragas}, {Vissapragada}, {Knutson},
  {Oklop{\v{c}}i{\'c}}, {Chachan}, {Greklek-McKeon}, {Dai}, {Tinyanont}, \&
  {Vasisht}}]{Paragas2021}
{Paragas}, K., {Vissapragada}, S., {Knutson}, H.~A., {et~al.} 2021, arXiv
  e-prints, arXiv:2102.08392.
\newblock \doarXiv{2102.08392}

\bibitem[{P\'erez \& Granger(2007)}]{PER-GRA:2007}
P\'erez, F., \& Granger, B.~E. 2007, Computing in Science and Engineering, 9,
  21, \dodoi{10.1109/MCSE.2007.53}

\bibitem[{{Piaulet} {et~al.}(2021){Piaulet}, {Benneke}, {Rubenzahl}, {Howard},
  {Lee}, {Thorngren}, {Angus}, {Peterson}, {Schlieder}, {Werner}, {Kreidberg},
  {Jaouni}, {Crossfield}, {Ciardi}, {Petigura}, {Livingston}, {Dressing},
  {Fulton}, {Beichman}, {Christiansen}, {Gorjian}, {Hardegree-Ullman}, {Krick},
  \& {Sinukoff}}]{Piaulet2021}
{Piaulet}, C., {Benneke}, B., {Rubenzahl}, R.~A., {et~al.} 2021, \aj, 161, 70,
  \dodoi{10.3847/1538-3881/abcd3c}

\bibitem[{{Pinto} \& {Galli}(2008)}]{Pinto2008}
{Pinto}, C., \& {Galli}, D. 2008, \aap, 484, 17,
  \dodoi{10.1051/0004-6361:20078819}

\bibitem[{{Roberge} \& {Dalgarno}(1982)}]{RobergeDalgarno1982}
{Roberge}, W., \& {Dalgarno}, A. 1982, \apj, 255, 489, \dodoi{10.1086/159849}

\bibitem[{{Rogers} {et~al.}(2021){Rogers}, {Gupta}, {Owen}, \&
  {Schlichting}}]{Rogers2021}
{Rogers}, J.~G., {Gupta}, A., {Owen}, J.~E., \& {Schlichting}, H.~E. 2021,
  arXiv e-prints, arXiv:2105.03443.
\newblock \doarXiv{2105.03443}

\bibitem[{{Schneiter} {et~al.}(2016){Schneiter}, {Esquivel}, {D'Angelo},
  {Vel{\'a}zquez}, {Raga}, \& {Costa}}]{Schneiter2016}
{Schneiter}, E.~M., {Esquivel}, A., {D'Angelo}, C.~S.~V., {et~al.} 2016,
  \mnras, 457, 1666, \dodoi{10.1093/mnras/stw076}

\bibitem[{{Shaikhislamov} {et~al.}(2021){Shaikhislamov}, {Khodachenko},
  {Lammer}, {Berezutsky}, {Miroshnichenko}, \&
  {Rumenskikh}}]{Shaikhislamov2021}
{Shaikhislamov}, I.~F., {Khodachenko}, M.~L., {Lammer}, H., {et~al.} 2021,
  \mnras, 500, 1404, \dodoi{10.1093/mnras/staa2367}

\bibitem[{{Shaikhislamov} {et~al.}(2016){Shaikhislamov}, {Khodachenko},
  {Lammer}, {Kislyakova}, {Fossati}, {Johnstone}, {Prokopov}, {Berezutsky},
  {Zakharov}, \& {Posukh}}]{Shaikhislamov2016}
---. 2016, \apj, 832, 173, \dodoi{10.3847/0004-637X/832/2/173}

\bibitem[{{Sing} {et~al.}(2019){Sing}, {Lavvas}, {Ballester}, {Lecavelier des
  Etangs}, {Marley}, {Nikolov}, {Ben-Jaffel}, {Bourrier}, {Buchhave}, {Deming},
  {Ehrenreich}, {Mikal-Evans}, {Kataria}, {Lewis}, {L{\'o}pez-Morales},
  {Garc{\'\i}a Mu{\~n}oz}, {Henry}, {Sanz-Forcada}, {Spake}, {Wakeford}, \&
  {PanCET Collaboration}}]{Sing2019}
{Sing}, D.~K., {Lavvas}, P., {Ballester}, G.~E., {et~al.} 2019, \aj, 158, 91,
  \dodoi{10.3847/1538-3881/ab2986}

\bibitem[{{Spake} {et~al.}(2021){Spake}, {Oklop{\v{c}}i{\'c}}, \&
  {Hillenbrand}}]{Spake2021}
{Spake}, J.~J., {Oklop{\v{c}}i{\'c}}, A., \& {Hillenbrand}, L.~A. 2021, arXiv
  e-prints, arXiv:2107.08999.
\newblock \doarXiv{2107.08999}

\bibitem[{{Spake} {et~al.}(2018){Spake}, {Sing}, {Evans}, {Oklop{\v{c}}i{\'c}},
  {Bourrier}, {Kreidberg}, {Rackham}, {Irwin}, {Ehrenreich}, {Wyttenbach},
  {Wakeford}, {Zhou}, {Chubb}, {Nikolov}, {Goyal}, {Henry}, {Williamson},
  {Blumenthal}, {Anderson}, {Hellier}, {Charbonneau}, {Udry}, \&
  {Madhusudhan}}]{Spake2018}
{Spake}, J.~J., {Sing}, D.~K., {Evans}, T.~M., {et~al.} 2018, \nat, 557, 68,
  \dodoi{10.1038/s41586-018-0067-5}

\bibitem[{{Stone} {et~al.}(2008){Stone}, {Gardiner}, {Teuben}, {Hawley}, \&
  {Simon}}]{2008ApJS..178..137S}
{Stone}, J.~M., {Gardiner}, T.~A., {Teuben}, P., {Hawley}, J.~F., \& {Simon},
  J.~B. 2008, \apjs, 178, 137, \dodoi{10.1086/588755}

\bibitem[{{Stone} \& {Proga}(2009)}]{StoneProga2009}
{Stone}, J.~M., \& {Proga}, D. 2009, \apj, 694, 205,
  \dodoi{10.1088/0004-637X/694/1/205}

\bibitem[{{Stone} {et~al.}(2020){Stone}, {Tomida}, {White}, \&
  {Felker}}]{2020ApJS..249....4S}
{Stone}, J.~M., {Tomida}, K., {White}, C.~J., \& {Felker}, K.~G. 2020, \apjs,
  249, 4, \dodoi{10.3847/1538-4365/ab929b}

\bibitem[{{Szab{\'o}} \& {Kiss}(2011)}]{SzaboKiss2011}
{Szab{\'o}}, G.~M., \& {Kiss}, L.~L. 2011, \apjl, 727, L44,
  \dodoi{10.1088/2041-8205/727/2/L44}

\bibitem[{Towns {et~al.}(2014)Towns, Cockerill, Dahan, Foster, Gaither,
  Grimshaw, Hazlewood, Lathrop, Lifka, Peterson, Roskies, Scott, \&
  Wilkins-Diehr}]{xsede}
Towns, J., Cockerill, T., Dahan, M., {et~al.} 2014, Computing in Science \&
  Engineering, 16, 62, \dodoi{10.1109/MCSE.2014.80}

\bibitem[{{Tripathi} {et~al.}(2015){Tripathi}, {Kratter}, {Murray-Clay}, \&
  {Krumholz}}]{Tripathi2015}
{Tripathi}, A., {Kratter}, K.~M., {Murray-Clay}, R.~A., \& {Krumholz}, M.~R.
  2015, \apj, 808, 173, \dodoi{10.1088/0004-637X/808/2/173}

\bibitem[{Van Der~Walt {et~al.}(2011)Van Der~Walt, Colbert, \&
  Varoquaux}]{van2011numpy}
Van Der~Walt, S., Colbert, S.~C., \& Varoquaux, G. 2011, Computing in Science
  \& Engineering, 13, 22

\bibitem[{{Van Eylen} {et~al.}(2018){Van Eylen}, {Agentoft}, {Lundkvist},
  {Kjeldsen}, {Owen}, {Fulton}, {Petigura}, \& {Snellen}}]{VanEylen2018}
{Van Eylen}, V., {Agentoft}, C., {Lundkvist}, M.~S., {et~al.} 2018, \mnras,
  479, 4786, \dodoi{10.1093/mnras/sty1783}

\bibitem[{{Vidal-Madjar} {et~al.}(2003){Vidal-Madjar}, {Lecavelier des Etangs},
  {D{\'e}sert}, {Ballester}, {Ferlet}, {H{\'e}brard}, \&
  {Mayor}}]{Vidal-Madjar2003}
{Vidal-Madjar}, A., {Lecavelier des Etangs}, A., {D{\'e}sert}, J.-M., {et~al.}
  2003, \nat, 422, 143, \dodoi{10.1038/nature01448}

\bibitem[{{Vidotto} \& {Cleary}(2020)}]{Vidotto2020}
{Vidotto}, A.~A., \& {Cleary}, A. 2020, \mnras, 494, 2417,
  \dodoi{10.1093/mnras/staa852}

\bibitem[{{Villarreal D'Angelo} {et~al.}(2018){Villarreal D'Angelo},
  {Esquivel}, {Schneiter}, \& {Sgr{\'o}}}]{Villarreal2018}
{Villarreal D'Angelo}, C., {Esquivel}, A., {Schneiter}, M., \& {Sgr{\'o}},
  M.~A. 2018, \mnras, 479, 3115, \dodoi{10.1093/mnras/sty1544}

\bibitem[{{Villarreal D'Angelo} {et~al.}(2014){Villarreal D'Angelo},
  {Schneiter}, {Costa}, {Vel{\'a}zquez}, {Raga}, \&
  {Esquivel}}]{Villarreal2014}
{Villarreal D'Angelo}, C., {Schneiter}, M., {Costa}, A., {et~al.} 2014, \mnras,
  438, 1654, \dodoi{10.1093/mnras/stt2303}

\bibitem[{{Villarreal D'Angelo} {et~al.}(2021){Villarreal D'Angelo}, {Vidotto},
  {Esquivel}, {Hazra}, \& {Youngblood}}]{Villarreal2021}
{Villarreal D'Angelo}, C., {Vidotto}, A.~A., {Esquivel}, A., {Hazra}, G., \&
  {Youngblood}, A. 2021, \mnras, 501, 4383, \dodoi{10.1093/mnras/staa3867}

\bibitem[{{Virtanen} {et~al.}(2020){Virtanen}, {Gommers}, {Oliphant},
  {Haberland}, {Reddy}, {Cournapeau}, {Burovski}, {Peterson}, {Weckesser},
  {Bright}, {van der Walt}, {Brett}, {Wilson}, {Jarrod Millman}, {Mayorov},
  {Nelson}, {Jones}, {Kern}, {Larson}, {Carey}, {Polat}, {Feng}, {Moore}, {Vand
  erPlas}, {Laxalde}, {Perktold}, {Cimrman}, {Henriksen}, {Quintero}, {Harris},
  {Archibald}, {Ribeiro}, {Pedregosa}, {van Mulbregt}, \&
  {Contributors}}]{2020SciPy-NMeth}
{Virtanen}, P., {Gommers}, R., {Oliphant}, T.~E., {et~al.} 2020, Nature
  Methods, \dodoi{https://doi.org/10.1038/s41592-019-0686-2}

\bibitem[{{Vissapragada} {et~al.}(2020){Vissapragada}, {Knutson}, {Jovanovic},
  {Harada}, {Oklop{\v{c}}i{\'c}}, {Eriksen}, {Mawet}, {Millar-Blanchaer},
  {Tinyanont}, \& {Vasisht}}]{Vissapragada2020}
{Vissapragada}, S., {Knutson}, H.~A., {Jovanovic}, N., {et~al.} 2020, \aj, 159,
  278, \dodoi{10.3847/1538-3881/ab8e34}

\bibitem[{{Wang} \& {Dai}(2018)}]{Wang2018}
{Wang}, L., \& {Dai}, F. 2018, \apj, 860, 175, \dodoi{10.3847/1538-4357/aac1c0}

\bibitem[{{Wang} \& {Dai}(2020{\natexlab{a}})}]{Wang2020a}
---. 2020{\natexlab{a}}, arXiv e-prints, arXiv:2101.00042.
\newblock \doarXiv{2101.00042}

\bibitem[{{Wang} \& {Dai}(2020{\natexlab{b}})}]{Wang2020b}
---. 2020{\natexlab{b}}, arXiv e-prints, arXiv:2101.00045.
\newblock \doarXiv{2101.00045}

\end{thebibliography}

\appendix

\section{Gas density and reaction rates}
\label{sec:reaction_rates}

\begin{figure*}[tbp]
\begin{center}
\includegraphics[width=0.95\textwidth]{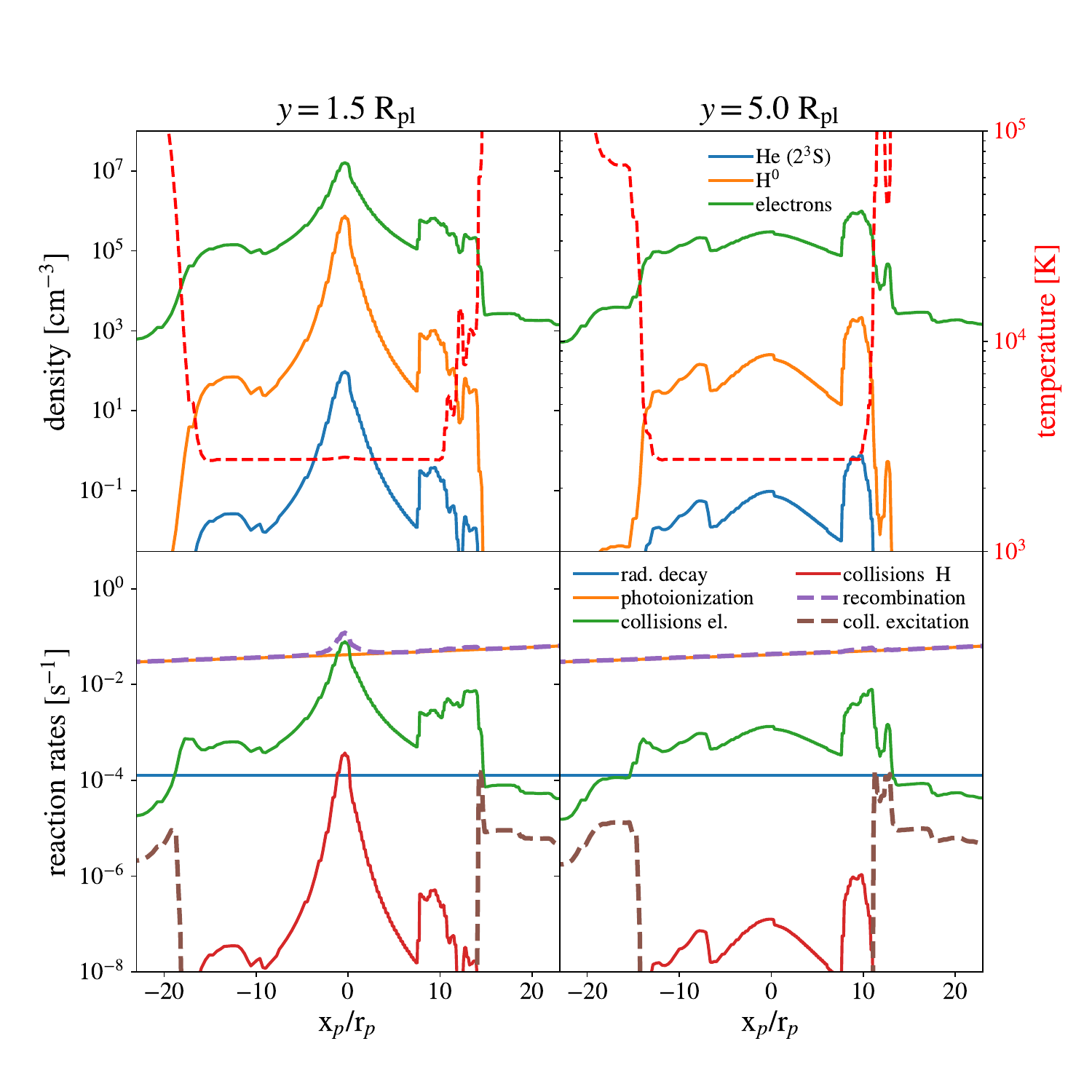}
\caption{Temperature and number density of metastable helium, neutral hydrogen, and free electrons along two rays parallel to the x-axis, with impact parameters $y=1.5$~R$_{\rm p}$ and $y=5.0$~R$_{\rm p}$ (top). The planetary and stellar winds are effectively isothermal, with (independent) temperatures set by two free parameters of our model, $\lambda_p$ and $\lambda_*$. The bottom panels show the rates of reactions responsible for populating (dashed lines) and depopulating (solid lines) the metastable state of helium. The shown rates correspond to different terms in Eq. (10) divided by the metastable helium fraction ($f_3$).  We note here that the phototoionization rate depends on the flux of the input spectrum as discussed in detail by \citet{Oklopcic2019} as well as the optical depth. Thus, relative rates are important, but their absolute value reflects the parameter choices of a given model or system. }
\label{fig:along_rays}
\end{center}
\end{figure*}

The top row of Figure \ref{fig:along_rays} shows the density and temperature structure along two rays close to the planet in the snapshot from model A. The helium opacity is significant in regions where the total gas number density is $\sim 10^6$ cm$^{-3}$ or greater. Collisional cross sections for most species relevant for our simulation (e.g. H$^+$--H, e--H, He--H$^+$ collisions) are in the range between 10$^{-15}$~cm$^2$ and 10$^{-13}$~cm$^2$ \citep{Pinto2008}. If we make a conservative estimate of 10$^{-15}$~cm$^2$, the mean free path of particles in that region is (at most) $\sim 10^9$~cm, which is $\sim 10$\% of the planet radius. Therefore, the mean free path is typically small compared to the relevant scale of the problem, which means that the planetary outflow probed by the helium line is in a collisional regime and the use of hydrodynamic simulations is appropriate.  A high frequency of collisions also means that the momentum imparted to metastable helium atoms by 1083~nm photons gets  distributed to the rest of the gas  upon each collision. This makes the effects of radiation pressure less important in the case of helium spectra, compared to the planetary wind morphology probed by the wings of the hydrogen Lyman-$\alpha$ line, which are sensitive to regions of much lower density, where collisional coupling may be ineffective.  

\section{Code Validation and Convergence}
\label{sec:numerical_param}
\subsection{Numerical Parameters}
\begin{figure*}[tbp]
\begin{center}
\includegraphics[width=0.99\textwidth]{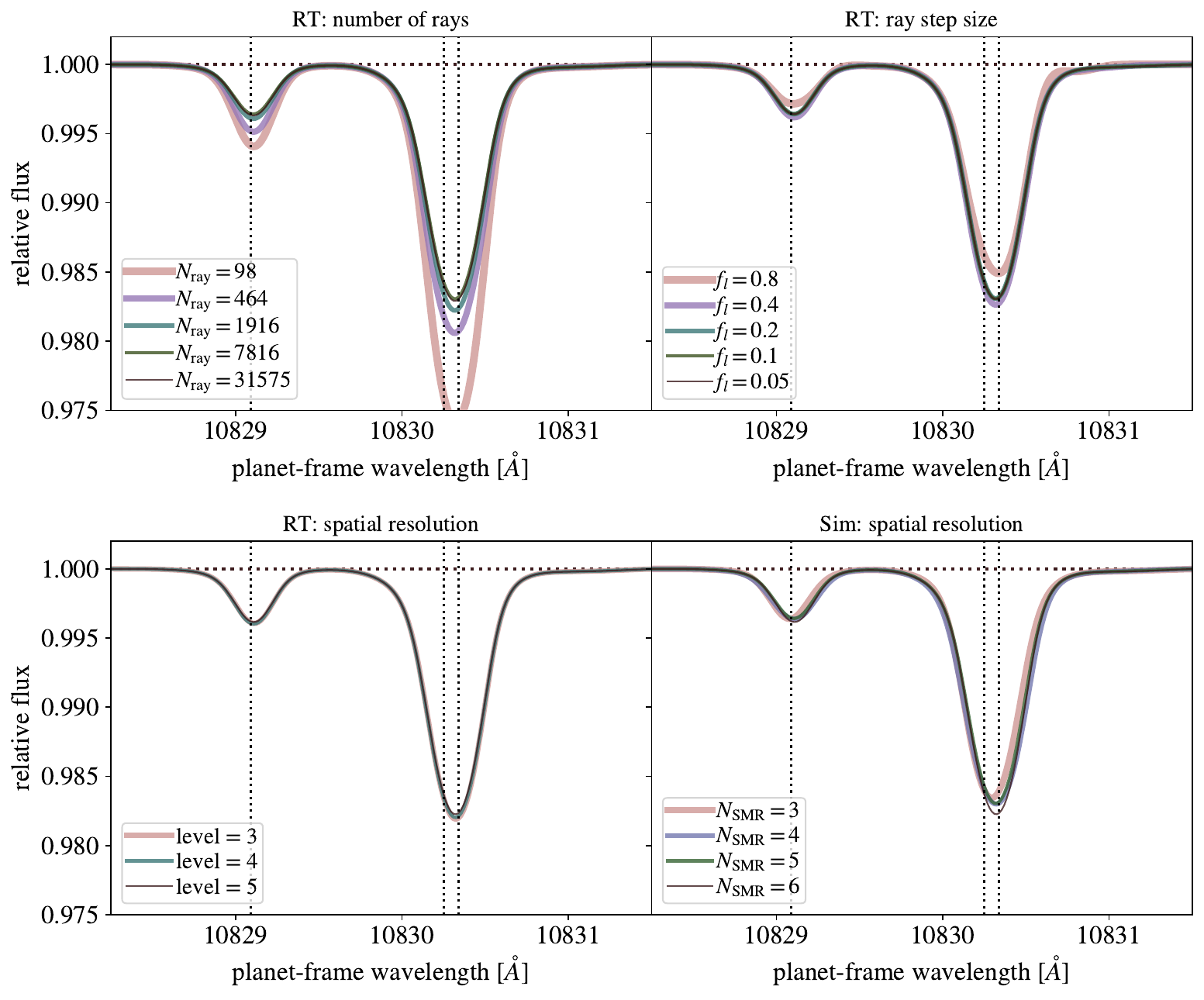}
\caption{Convergence studies showing spectra computed for the parameters of model A, with various numerical choices. Each spectrum is plotted at a phase of 0.0095, near ingress. The first panel varies the spatial resolution of the simulation itself in the box surrounding the planet via the number of levels of static mesh refinement, $N_{\rm SMR}$ (models A3, A4, A, and A6 in Table \ref{simtable}, respectively). The second panel focuses on model A, and varies the spatial resolution during the radiative transfer post-processing step. The third panel varies the number of points along radiative transfer rays cast through the domain. For context, we note that these spectra have narrower thermal width, and thus a more-distinct triplet feature than models tailored to the higher-temperature outflow of WASP-107b \citep[e.g.][]{Wang2020b,Khodachenko2021}.  }
\label{fig:convergence}
\end{center}
\end{figure*}

In this section we document several tests intended to validate aspects of our numerical hydrodynamics and post-processing schemes. We focus our attention on variations of model A and its radiative transfer post processing. Each model is plotted at a phase of 0.0095, which is near ingress of the optical transit.  

The surface of the star is sampled by rays arranged in impact parameter, $b$, and angle $\theta$, relative to the planet. We allocate the faces between rings according to a power-law in $b$, with $b^{0.75}$ proving to be an effective balance of higher sampling density near the planet balanced with sufficient sampling at larger impact parameters. The ring width is $db$, and the each ring is divided into an integer number of zones in $\theta$, such that $d\theta \approx db/b$. The centers of these zones are computed from an area-weighted average. We cast rays from where the zone centers intersect the stellar surface through the computational domain. The step size $dl$ along these rays is chosen such that $dl \approx f_l d_p$, where $d_p$ is the three-dimensional distance from the planet. Thus, the sampling length is smaller closer to the planet where there are more compact hydrodynamic structures to resolve. 

In the first panel of Figure \ref{fig:convergence}, we examine the convergence of synthetic spectra with different total numbers of rays. We find that our models converge satisfactorily to a solution with more than a few thousand rays. Next, we examine the step size along rays, as modulated by the parameter $f_l$. Here we find that as long as the dimensionless step size is smaller than 0.4, our final spectra are very similar.

In the lower panels, we examine the effects of spatial resolution, first in our radiative transfer post-processing, then in our hydrodynamic simulations themselves. Resolution is controlled by the number of static mesh refinement levels in the box that extends 10 planet radii around the planet. The lower left panel demonstrates minimal differences between the spectra from model A6 downsampled to post-process at maximum refinement level 3, 4, or 5, decreasing the effective spatial resolution during the post-processing but not during the hydrodynamic evolution itself. This downsampling is valualble because the ray-tracing is memory intensive, and each downsampling reduces the memory usage by a factor of 8. We find that the various degrees of downsampling are nearly indistinguishable, which agrees with our results for $f_l$ that small scale features do not dominate the emergent spectra. Finally, we find that our hydrodynamic models with maximal refinements of 3, 4, 5, and 6 levels around the planet all achieve similar (though not completely identical) results, with the line depth being slightly larger in the most-finely resolved case. The relevant models here are A3, A4, A, and A6, which are tabulated in Table \ref{simtable}. From Table  \ref{simtable} we note that these models generate similar, but not identical, time-averaged outflow rates from the star and planet. In model A3, the resolution element is $\delta r \approx 0.495R_p$, or about half of the planet radius. In model A4, $\delta r \approx 0.248R_p$, in model A, $\delta r \approx 0.124R_p$, and in model A6, $\delta r \approx 0.062R_p$. These represent approximately 2, 4, 8, and 16 zones across the planet radius in models A3, A4, A, and A6, respectively. We note that it is important that the planet radius is close to divisible by the resolution element given that our mesh is approximately Cartesian in the vicinity of the planet because the origin of the spherical polar geometry is at the star (otherwise the planet's numerical size would change with resolution choice).

From these studies we adopt our fiducial parameters of running the models at $N_{\rm SMR}=5$, and post-processing the radiative transfer downsampled by a factor of two at level 4, we choose $f_l=0.05$ in case of smaller-scale features than are present in case A, and 100 radial bins of sampling points, which corresponds to the $N_{\rm ray}=7816$ case at a phase of 0.0095, and $N_{\rm ray} \approx 32,000$ near mid-transit.

\section{Data and Software Availability}

We make the software and data needed to reproduce the results and analysis in this paper public in parallel with this article. Hydrodynamic model snapshots (Athena++ hdf5 format), synthetic spectra (ascii format), and three dimensional post-processed species number densities (hdf5 format) are available on Zenodo at \dataset[10.5281/zenodo.5750747]{https://zenodo.org/record/5750747}.  The radiative transfer post-processing software that we apply, along with the code that reproduces our figures on the basis of the post-processed data is available on \dataset[Github]{https://github.com/morganemacleod/pw_sw_materials} and Zenodo at \dataset[10.5281/zenodo.5750777]{https://zenodo.org/record/5750777}.

\end{document}